\documentclass[aps,prd,onecolumn,amsmath,amssymb,preprint,nofootinbib,tightenlines,superscriptaddress,longbibliography]{revtex4-1}
\pdfoutput=1
\usepackage{graphicx,subfigure}
\usepackage{dcolumn}
\usepackage{bm}
\usepackage{mathrsfs}
\def\ba{\begin{eqnarray}}
\def\ea{\end{eqnarray}}
\def\beq{\begin{eqnarray}}
\def\eeq{\end{eqnarray}}
\def\be{\begin{eqnarray}}
\def\ee{\end{eqnarray}}
\def\ben{\begin{equation} \nonumber}
\def\een{\end{equation}}
\def\baray{\begin{eqnarray*}}
\def\earay{\end{eqnarray*}}

\def\dh{d_h}
\def\fr{f_r}
\def\fm{f_m}
\def\nr{n_r}
\def\nm{n_m}

\def\teq{t_{\rm eq}}

\def\zeq{z_{\rm eq}}
\def\aeq{a_{\rm eq}}

\def\Alpha{\mbox{\Large $\alpha$}}
\def\Alphaeq{\mbox{\Large $\alpha$}_{\rm eq}}
\def\alphaeq{\mbox{\normalsize $\alpha$}_{\rm eq}}
\def\meq{M_{\rm eq}}

\def\nrecent{n_{\rm recent}}
\def\tnr{t_{\rm nr}}
\def\amin{{\alpha_{\rm min}}}
\def\xmin{{x_{\rm min}}}

\begin{document}

\title{The number of cosmic string loops}

\author{Jose J. Blanco-Pillado}
\affiliation{Department of Theoretical Physics, University of the
  Basque Country, 48080, Bilbao, Spain}
\affiliation{IKERBASQUE, Basque Foundation for Science,  48011, Bilbao, Spain}
\affiliation{Institute of Cosmology, Department of Physics and Astronomy, 
Tufts University, Medford, MA  02155, USA}
\author{Ken D. Olum}
\author{Benjamin Shlaer}
\affiliation{Institute of Cosmology, Department of Physics and Astronomy, 
Tufts University, Medford, MA  02155, USA}

\begin{abstract}
Using recent simulation results, we provide the mass and speed
spectrum of cosmic string loops.  This is the quantity of primary
interest for many phenomenological signatures of cosmic strings, and
it can be accurately predicted using recently acquired detailed
knowledge of the loop production function.  We emphasize that
gravitational smoothing of long strings does not play any role in
determining the total number of existing loops.  We derive a bound on
the string tension imposed by recent constraints on the stochastic
gravitational wave background from pulsar timing arrays, finding $G\mu
\leq 2.8\times 10^{-9}$.  We also provide a derivation of the
Boltzmann equation for cosmic string loops in the language of
differential forms.
\end{abstract}

\maketitle

\tableofcontents

\section{Introduction}

A longstanding problem in the theory of cosmic strings is how to
predict the distribution of loops that exist at any given time.  This
distribution provides the input for calculations of many observable
effects.  For example, stochastic gravitational wave signals
\cite{Vilenkin:1981bx,Hogan:1984is,Vachaspati:1984gt,Accetta:1988bg,
  Bennett:1990ry,Caldwell:1991jj,Siemens:2006yp,DePies:2007bm,Olmez:2010bi,Sanidas:2012ee,Sanidas:2012tf,Binetruy:2012ze,Kuroyanagi:2012wm,
  Kuroyanagi:2012jf} as well as bursts
\cite{Damour:2000wa,Damour:2001bk,Damour:2004kw,
  Siemens:2006vk,Hogan:2006we,Regimbau:2011bm} are proportional to the
number of loops above a critical size dependent on the frequency band
of the detector.  The ionization history of the universe is affected
by loops through early star formation
\cite{Olum:2006at,Shlaer:2012rj}, which is seeded by the large, slow
loops.  Neutral hydrogen overdensities caused by loops
\cite{Pagano:2012cx, Tashiro:2013aaa} create bright spots in
high-redshift 21cm surveys, and dark matter overdensities
\cite{Berezinsky:2011wf} have enhanced decay signatures.  Microlensing
of stars may occur from loops slow enough to be captured by galaxy
halos, even for very low tensions \cite{Chernoff:2007pd,
  Chernoff:2009tp,Pshirkov:2009vb}.  Magnetogenesis on galactic scales
\cite{Battefeld:2007qn} may be aided by large cosmic string loops.
Cosmic rays from ordinary \cite{Brandenberger:1986vj,MacGibbon:1989kk,
  Mohazzab:1993ah,BlancoPillado:1998bv,Olum:1998ag,Berezinsky:2011cp}
and superconducting cosmic strings
\cite{Hill:1986mn,Berezinsky:2009xf,
  Vachaspati:2008su,Cai:2011bi,Cai:2012zd} are emitted via cusps and
kinks on loops. Additional couplings of the string to other degrees of
freedom could also lead to other forms of radiation by loops
\cite{Srednicki:1986xg, Damour:1996pv,Peloso:2002rx,Babichev:2005qd,
Vachaspati:2009kq,Dufaux:2012np,Lunardini:2012ct}.

Our goal here is to provide a definitive description of the number
density of loops of a given size and velocity, based on recent
simulations \cite{BlancoPillado:2011dq, BlancoPillado:2010sy}.
Simulations give us the rate of production of the loops over the
timescale of the simulations.  We then extrapolate these results and
embed them in a cosmological context.  While the simulations have not
yet determined the production rate of loops that are very small
relative to the horizon size, the production of larger loops is
reasonably well agreed upon.  As we will show, the uncertainty about
the production of tiny loops is of no consequence for any known
observable effect, because small loops produced recently are dwarfed
by loops of the same (current) size produced long ago, when the string
network (and the universe) were much more dense.

The reason that it is possible to extrapolate from simulations, which
can cover only a few orders of magnitude in the growth of the
universe, to cosmological phenomenology is the {\em scaling} nature of
cosmic string networks;  there is only one kinematic scale, which we take to 
be the horizon distance, and all other length scales, such as
the Hubble length, are proportional to the horizon distance.  By
understanding the self-similar dynamics numerically, we can
extrapolate the network behavior through all cosmological epochs.

Scaling of the loop distribution means that the number of loops of a
given fraction of the horizon size contained in a single horizon
volume should not change with time, once the scaling distribution is
established.  The earliest work \cite{Albrecht:1984xv,Albrecht:1989mk}
seemed to indicate a scaling distribution, but subsequent numerical
simulations
\cite{Bennett:1987vf,Bennett:1989ak,Bennett:1989yp,Allen:1990tv} found
instead that most energy went to loops at the minimum resolution
allowed in the simulations, rather than scaling with the horizon size.
Extrapolating those results in a cosmological context, one would
conclude that loops are always produced at a scale determined by
gravitational back-reaction, since this is smallest relevant scale.

The question then arose \cite{Vanchurin:2005yb,Ringeval:2005kr}
whether the simulation-resolution loops represented a real feature of
a cosmic string network or were rather an artifact of the initial
conditions.  Indeed, more recent simulations have determined that a
significant fraction of loops are produced at scales roughly a few
orders of magnitude below the horizon size, after a transient initial
regime.  In Ringeval, Sakellariadou, and Bouchet
\cite{Ringeval:2005kr} (hereafter RSB), a scaling sub-population of
large loops was first shown to exist over a range of sizes within a
few orders of magnitude of the horizon scale.  Subsequently,
Refs.~\cite{Vanchurin:2005pa,Olum:2006ix,BlancoPillado:2011dq} found
scaling in the rate of production of loops, rather than the
distribution of existing loops.  Although slightly different
definitions\footnote{Refs.~\cite{Olum:2006ix,BlancoPillado:2011dq}
  only considered loops on non-self-intersecting trajectories, while
  RSB considered any closed string shorter than the horizon to be a
  loop.} were used, the phenomenological results are in agreement,
namely that the spectrum of loops in horizon units is set by a scale
of order unity, rather than the gravitational smoothing scale.  The
most recent simulation \cite{BlancoPillado:2011dq} found scaling of
the loop production function over five orders of magnitude in loop
size, which is enough to determine the loop distribution to within a
few percent.  The agreement between \cite{BlancoPillado:2011dq} and
\cite{Ringeval:2005kr} is rather good in both the radiation and matter
eras, as our figures will show.

Reference~\cite{Lorenz:2010sm} is not directly comparable with this
paper, since an analytic model of loop production was used to bridge
the gap between simulation data for horizon-scale loops, and the
gravitational back-reaction scale.  Their analytic model assumes most
loops were produced at the gravitational back-reaction scale, and
hence differs quite significantly from our simulated loop production
function.  In particular, we find that the vast majority of loops
existing at any given time were produced with a size within a few
orders of magnitude of the horizon size at the time of production.

An important consideration which has not been widely appreciated is
the speed of loops emitted from the network \cite{Polchinski:2006ee}.
Small loops are created with ultra-relativistic speeds, and thus lose
most of their energy to redshifting.  Except for loops so small that
they decay within a few Hubble times, it is better to classify loops
by their rest mass (i.e., the total energy of a loop viewed in the
frame where the center of mass is at rest).  We find that a negligible
fraction of mass is injected into loops well below the horizon size.
Thus, it is loops produced within a few orders of magnitude of the
horizon size that account for most of the loops of any given size that
exist at any given time.

Large loops are emitted with velocities on the order of 0.3, where 
we assume the speed of light is unity throughout.  In terms
of Lorentz boost, this is of little consequence, but if one is
concerned with seeding structure formation \cite{Shlaer:2012rj} or
capture of small loops in galaxies \cite{Chernoff:2009tp}, even
non-relativistic velocities are important.

\section{Cosmic string loop production and decay}

\subsection{Scaling of loops}

Simulations have revealed that the long strings in a cosmic string
network obey a scaling solution \cite{Albrecht:1984xv,Bennett:1987vf,
Albrecht:1989mk,Bennett:1989ak,Bennett:1989yp,Allen:1990tv,
Vanchurin:2005pa,Ringeval:2005kr,Martins:2005es,Olum:2006ix,
BlancoPillado:2011dq}.  We will describe all scaling
quantities using the horizon distance $\dh$, which in a radiation
dominated or matter dominated universe is related to cosmic time $t$
by $\dh = 2t$ or $\dh = 3t$, respectively.  We will specify the comoving loop
production function as a function\footnote{Such quantities can more
  elegantly be described in the language of differential forms.  See
  Appendix \ref{sec:appendix-forms}.} $f(t,m,p)$, where $t$
is the time of production, $m$ the rest mass of the loop, and $p$ is
the momentum per unit mass, i.e., $p=v\gamma = v/\sqrt{1-v^2}$, where
$v$ is the center-of-mass velocity of the loop and $\gamma$ is its
Lorentz boost.  We will sometimes refer to $p$ as just the momentum.
Then $f(t,m,p) dt\,dm\,dp$ gives the number of loops produced per comoving volume in time
interval $dt$ with rest mass between $m$ and $m + dm$ and momentum per
unit mass between $p$ and $p + dp$.

We define true scaling when not just the normalization, but also the
shape of the spectrum of power flowing into loops becomes
time-independent when $m$ is expressed in units of $\mu\dh$.  Thus we
define a scaling measure of the loop mass
\be
\alpha = \frac{m}{\dh\mu},
\ee
and a scaling loop production function $f(\alpha, p)$, so that
$f(\alpha, p)d\alpha\,dp$ is essentially the number of loops produced in volume
$\dh^3$ in time $\dh$ with $\alpha$ and $p$ in a range of size $d\alpha$ and $dp$.
More precisely,
\ba
\frac{f(\alpha,p)}{\dh^4} = \frac{f(t,m,p)}{a^3}\frac{\partial m}{\partial \alpha},
\ea
where $\partial m/\partial \alpha = \mu \dh$ is the Jacobian determinant for changing coordinates
from $m$ to $\alpha$. 
Hence,
\be
f(\alpha,p) = \frac{\mu\dh^5}{a^3} f(t,m,p).
\ee
Numerical evidence now exists that the loop production power will
eventually scale\footnote{In fact, numerical simulations to date have
  only shown that large loop power scales, and that the majority of
  power is flowing into small loops not yet proven to scale.  We will
  see that enough large loop production has been found that the fate
  of these small loops is immaterial for the purposes of calculating the number
  density of loops $n$, although we believe small loop power does eventually scale,
  even in the absence of gravitational backreaction.  We no longer
  \cite{BlancoPillado:2011dq} believe that the large loop portion of
  $f$ will be significantly affected by this.}.

We characterize the distribution of cosmic string loops at time $t$ by
$n(t,m,p)dm\,dp$, the comoving number density of loops of rest mass between $m$
and $m + dm$ and momentum per unit mass between $p$ and $p + dp$.  The
scaling number density distribution is given by
$n(\alpha,p)d\alpha\,dp$, the number of loops in volume $\dh^3$ whose
$\alpha$ and $p$ are in ranges of size $d\alpha$ and $dp$.  The two
functions are related by
\ba\label{eq:nofalpha}
n(\alpha,p) = \frac{\mu\dh^4}{a^3} n(t,m,p).
\ea

The loop distribution can be determined by integrating the loop
production function.  (We assume that strings were formed early enough
that no loop present at the time of string formation could survive to any
time of interest.)  The number of loops in a given comoving volume at
time $t$ with $m$ and $p$ in given ranges is just the total number of
loops produced at all earlier times whose mass and momentum will,
after evolving to time $t$, fall in the given ranges.  Thus
\be\label{eq:simplen}
n(t,m,p) = 
\int_{0}^t f(t',M',P') \frac{\partial M'}{\partial m}\frac{\partial P'}{\partial p}\,dt' ,
\ee
where $M'$ and $P'$ are the mass and momentum at time $t'$ of a loop
which will eventually have mass $m$ and momentum $p$ at time
$t$.\footnote{The quantities $M'$ and $P'$ depend on the time
  of production, $t'$, and also on the loop parameters of interest,
  $m$ and $p$, and the time $t$ at which it has those parameters.  We
  can write them as functions $M' = M(t';t,m,p)$ and $P' =
  P(t';t,m,p)$.  These functions define the {\em flow}, which,
  along with Eq.~(\ref{eq:simplen}), is derived more formally in
  Appendix \ref{sec:appendix-forms}.  Here and below, we use capital
  symbols to indicate solutions to the flow, and primed symbols
  indicate that the suppressed time argument is $t'$, rather than $t$.}

Loops oscillate and decay by emission of gravitational radiation, and
the momentum of a loop decreases with the expansion of the universe.
We will neglect the change of the momentum due to gravitational wave
emission, the so-called ``rocket effect'', and consider only
redshifting, so the momentum is just inversely proportional to the
scale factor,
\be\label{eq:p}
P'=p\frac{a}{a'},
\ee
and so $\partial P'/\partial p = a/a'$. 
A more careful treatment is
necessary if one wishes to consider the effects of very slow loops,
e.g., gravitational clustering \cite{Chernoff:2007pd, Chernoff:2009tp}.

The rate of gravitational radiation from a loop does not depend on its
size.  Thus a loop that was slightly more massive than another at
production will be the same amount more massive today, and thus
\be\label{eq:mmp}
\frac{\partial M'}{\partial m} = 1.
\ee
Putting Eqs.~(\ref{eq:p}) and (\ref{eq:mmp}) in Eq.~(\ref{eq:simplen})
we find
\be\label{eq:simplen2}
n(t,m,p) = \int_{0}^t f(t',M',P')\frac{a}{a'}\,dt' ,
\ee
which in scaling coordinates becomes 
\be\label{eq:nrm1}
n(\alpha,p)=\int_0^t\frac{a'^2}{a^2}\frac{\dh^4}{\dh'^5}f(\Alpha',P') dt',
\ee
where $\Alpha' = M'/(\dh'\mu)$.

We can change variables and integrate over the scaling mass of the
loop at production, rather than the time of production to get
\be
n(\alpha,p) = \int^\alpha_\infty \frac{a'^2}{a^2}\frac{\dh^4}{\dh'^5}
f\left(\alpha',p \frac{a}{a'}\right) \frac{\partial t'}{\partial
  \alpha'}d\alpha'.
\ee

For most purposes, as we will discuss, it is sufficient to consider
only nonrelativistic center-of-mass speeds.  In that case, the change
of mass is given by the gravitational radiation power $\Gamma G \mu^2$
without accounting for time dilation, so
\be\label{eq:nrm}
M'\approx m + \Gamma G \mu^2 \left(t - t'\right),
\ee
where $\Gamma$ is a number of order $50$--$100$.
The exact flow of the mass is given in Appendix \ref{sec:exact-flow}.

In a radiation or matter dominated universe, we can write the scale factor
$a\propto t^\nu$, and thus $d_h = t/(1-\nu)$, with $\nu = 1/2$ for
radiation and $\nu = 2/3$ for matter.  Thus using Eq.~(\ref{eq:nrm})
we find
\be
\frac{t'}{t} = \frac{\alpha + (1-\nu)\Gamma G\mu}
{\alpha' + (1-\nu)\Gamma G\mu},
\ee
and so
\be\label{eq:nv}
n(\alpha, p) = \frac{(1-\nu)\int_\alpha^\infty
\left(\alpha' + (1-\nu)\Gamma G\mu\right)^{3-2\nu}f\left(\alpha', p\frac{a}{a'}\right)d\alpha'}
{\left(\alpha + (1-\nu)\Gamma G\mu\right)^{4-2\nu}},
\ee
with
\be
\frac{a}{a'}p = \left(\frac{t}{t'}\right)^\nu p
= \left(\frac{\alpha' + (1-\nu)\Gamma G\mu}
{\alpha + (1-\nu)\Gamma G\mu}\right)^\nu p.
\ee
In most cases we will not be very sensitive to the precise speed of
the loops, so we can integrate over $p$ to get the density of loops
without regard to momentum,
\be\label{eq:n}
n(\alpha) = \frac{(1-\nu)\int_\alpha^\infty
\left(\alpha' + (1-\nu)\Gamma G\mu\right)^{3-3\nu}f\left(\alpha'\right)d\alpha'}
{\left(\alpha + (1-\nu)\Gamma G\mu\right)^{4-3\nu}},
\ee
where we have defined $f(\alpha) = \int_0^{\infty}{ f(\alpha,p) dp}$.

\subsection{Radiation era}
\begin{figure}
   \centering
   \includegraphics[width=6.0in]{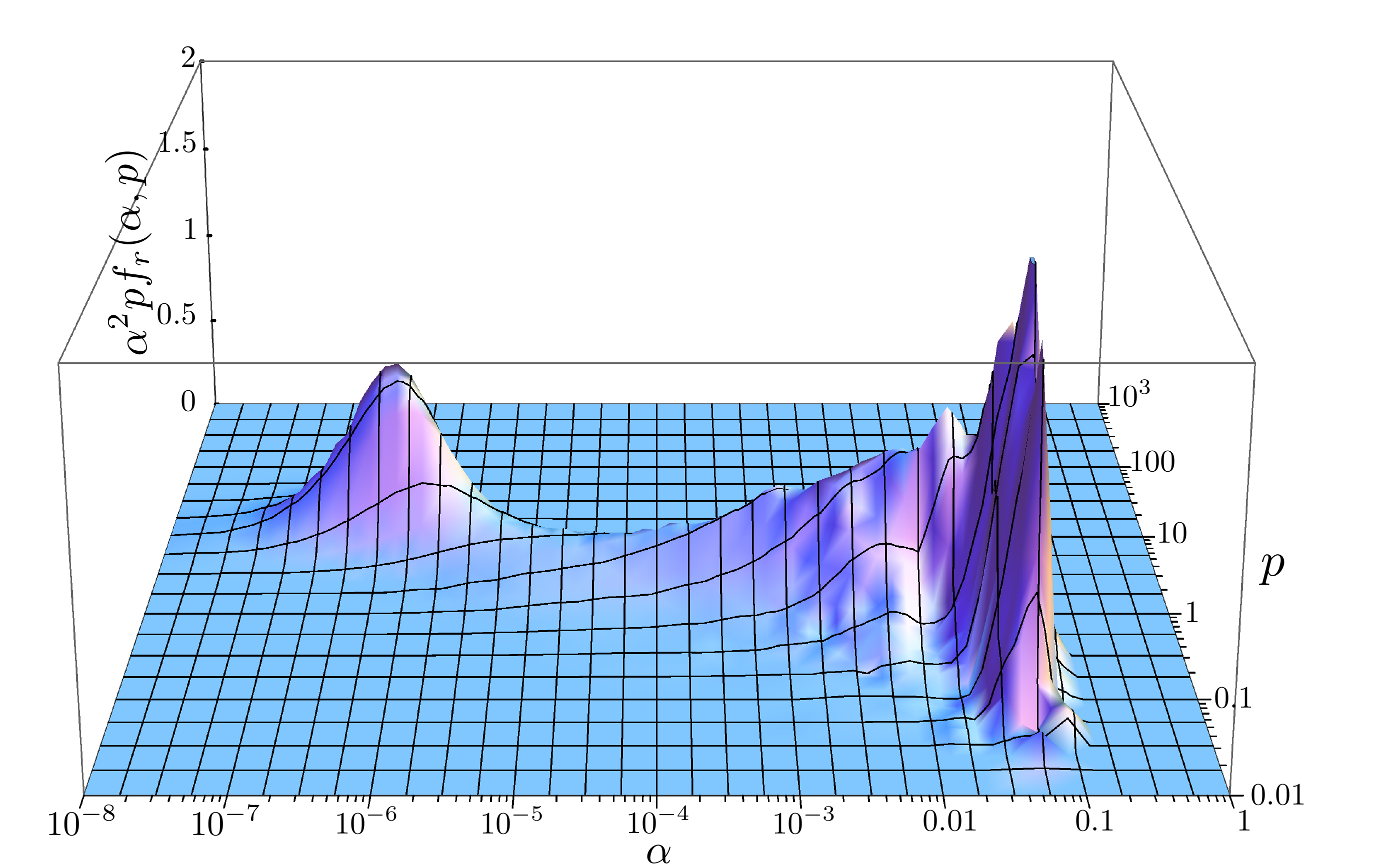} 
   \caption{The scaling rate of mass flowing into cosmic string loops
     in a radiation dominated universe simulation.  The spectrum is
     given in logarithmic bins of scaling mass $\alpha = m/(\mu \dh)$
     and momentum/mass $p=v/\sqrt{1-v^2}$.  The vertical axis,
     $\alpha^2 p f(\alpha,p)$, is chosen so that the volume under the
     surface gives the total mass flow.  (This effect is
       described more fully in Appendix \ref{sec:appendix-forms}.)  The
     peak at $\alpha \sim 10^{-6}$ is non-scaling, an artifact of the
     resolution of initial conditions. This error is a subdominant
     total mass fraction flowing into loops, but as visible in
     Ref.~\cite{BlancoPillado:2011dq}, it is a dominant energy
     fraction, due to very large speeds of small loops.  We will show
     that this artifact has no effect on the spectrum of loops
     $n(\alpha)$.}
   \label{fig:afrad}
\end{figure}

We show in Fig.~\ref{fig:afrad}
the rate of mass produced in cosmic string loops from 8 radiation-era
simulations in a box of size 1500 initial correlation lengths with
starting conformal time 6, ending time 1506 and thus dynamic range
251.  For details on the simulation procedure see
Refs.~\cite{BlancoPillado:2011dq,BlancoPillado:2010sy}.  For large
loops, the Lorentz boost is typically small, around 1.1, but for very
small loops it is quite high.  (This happens because the string is
sufficiently smooth at small scales that the only way for small loops
to be emitted is for the underlying string to be highly contracted and
boosted.)  As a result, the non-scaling peak in mass production at
small $\alpha$ (scaling mass) is much smaller than the non-scaling
energy production peak at small $x$ (scaling energy) shown in
Ref.~\cite{BlancoPillado:2011dq}.

\begin{figure}
   \centering
   \hspace{-.2in}
   \subfigure{
     \includegraphics[width=3.1in]{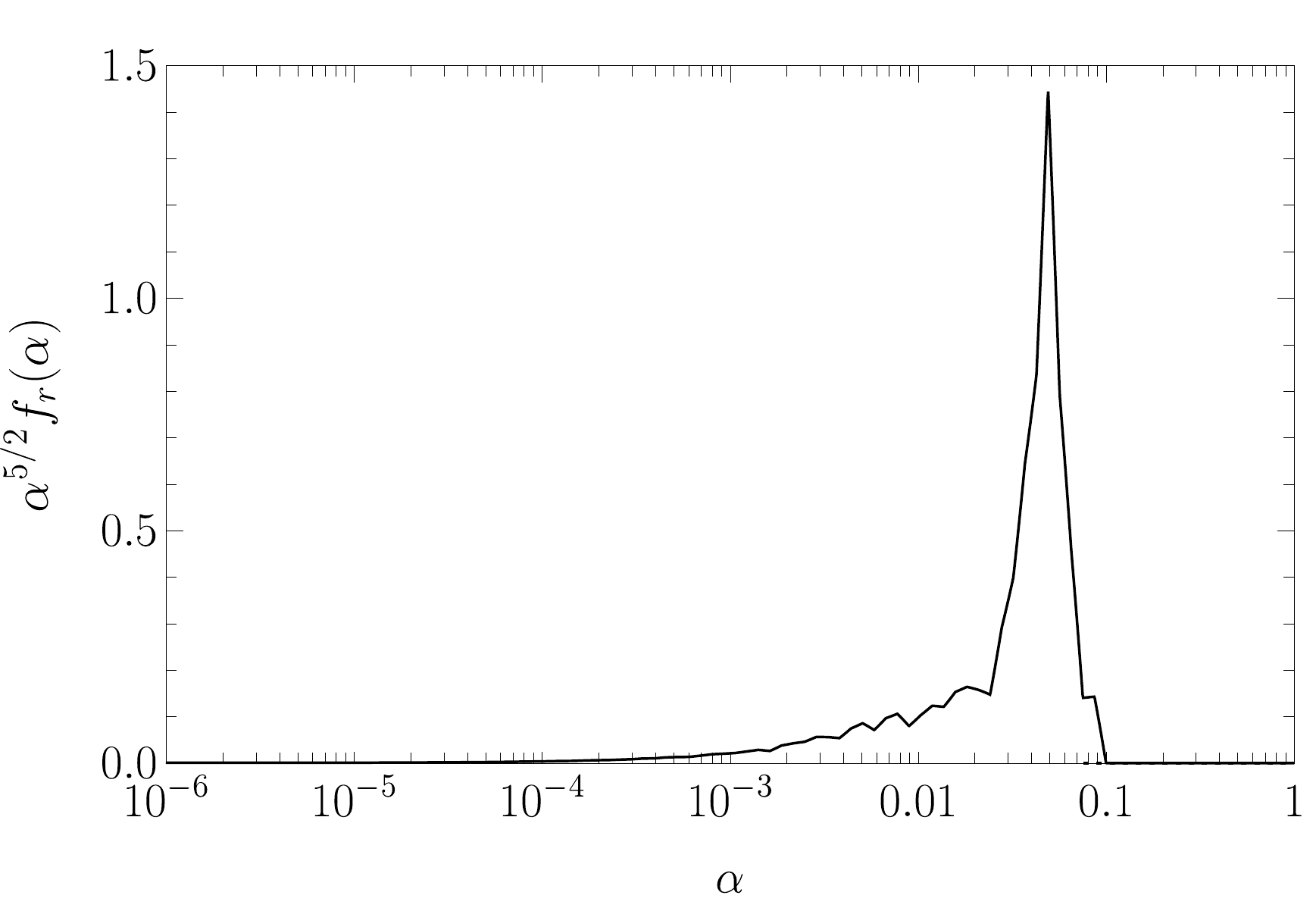} 
   }
   \subfigure{
   \includegraphics[width=3.2in]{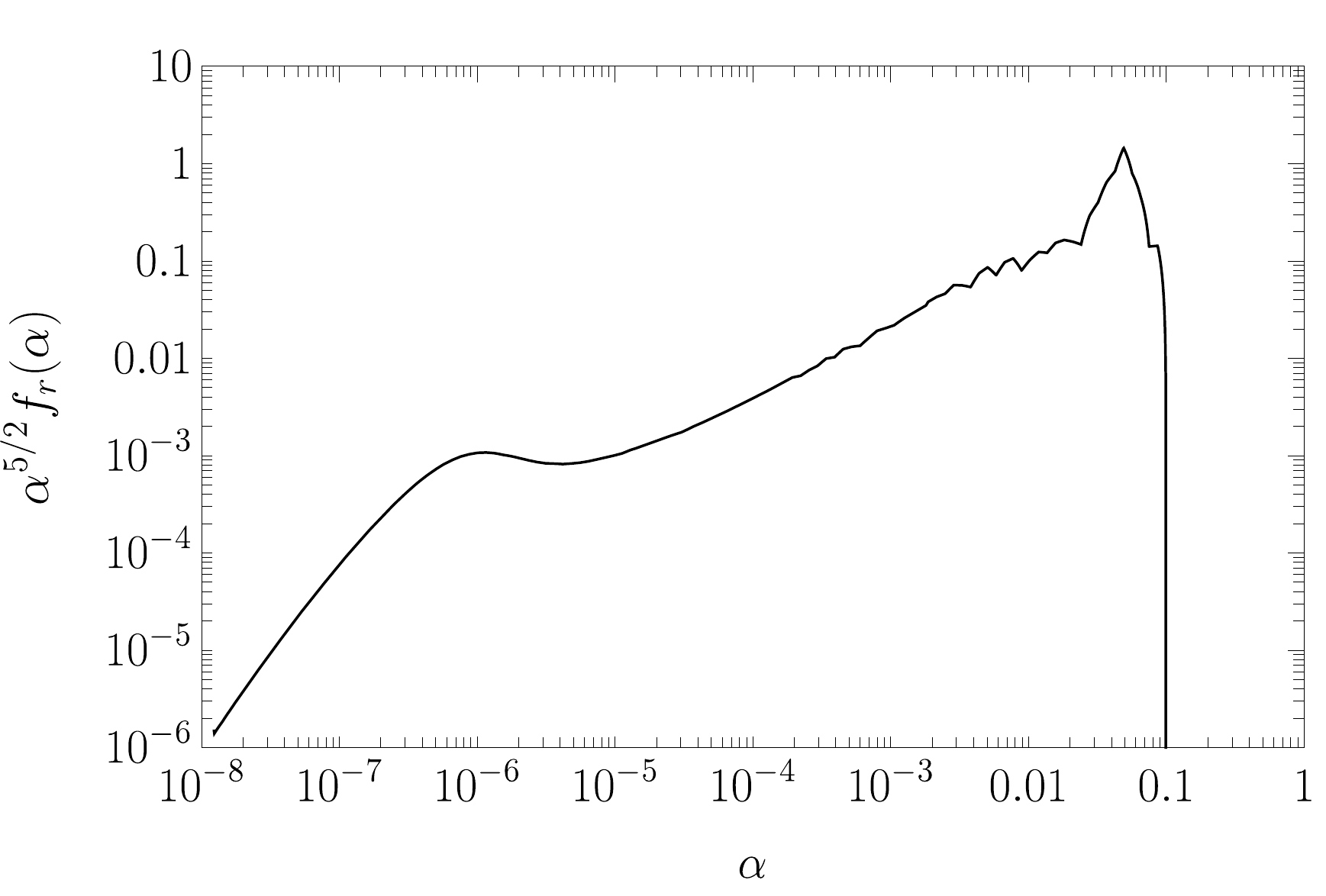} 
     }
   \caption{The loop production rate $\fr(\alpha)$ scaled by
     $\alpha^{5/2}$, as a function of $\alpha$ on a logarithmic scale.
     Because $\int \alpha^{3/2}\fr d\alpha = \int \alpha^{5/2}\fr
     d\ln\alpha$, the area under the curve on the left gives the
     contribution of each region of $\alpha$ to Eq.~(\ref{eq:nr}).
     The right-hand panel is the same with a logarithmic vertical
     scale.  Notice that the non-scaling peak at $\alpha \sim 10^{-6}$
     contributes a negligible fraction of loops.}
   \label{fig:a3-2frad}
\end{figure}

We can disregard momenta and find the distribution of loops by
mass alone using Eq.~(\ref{eq:n}) with $\nu = 1/2$,
\be\label{eq:nr}
\nr(\alpha) = \frac{\int_\alpha^\infty (\alpha' + \Gamma G\mu/2)^{3/2}
\fr(\alpha')d \alpha'}{2(\alpha + \Gamma G\mu/2)^{5/2}}.
\ee
Let us look first at the numerator.  If gravitational effects can be
neglected, it is just
$\int_\alpha^\infty{\alpha'}^{3/2} \fr(\alpha')d \alpha'$.
We plot the integrand in Fig.~\ref{fig:a3-2frad}.
There is an extra half power of $\alpha'$ relative to
Fig.~\ref{fig:afrad}.  It appears because the energy density in loops is only
diluted as $1/a^3$, whereas the network is putting energy into loops
as $1/a^4$.
Thus the network produces many more loops at early
times, when the network density is higher, so large loops from early
times are more numerous than loops of identical physical size produced
later on, despite dilution from the intervening expansion.  As a
result of this extra half power, the dominant contribution to loops of
any scaling mass $\alpha$ comes from production of large loops, and
the peak at small scales gives no significant contribution; for any
realistic $G\mu$, loops with $\alpha'\sim \Gamma G \mu$ give no
substantial contribution, as shown in Fig.~\ref{fig:a3-2frad}.  Thus,
even though we do not know the eventual fate of the small-scale peak,
that uncertainty does not yield any important uncertainty in the
resulting $n(\alpha)$.

What about $\alpha' < \Gamma G \mu$?  
In this range the coefficient of $f_r(\alpha')$ does not decrease 
below $(\Gamma G \mu)^{3/2}$ as $\alpha'$ gets smaller. 
Assuming excitations on strings with wavelengths less
than $\Gamma G\mu t$ are strongly suppressed, $f(\alpha')$ drops
rapidly for $\alpha' < \Gamma G \mu$, so this part of the integral does not
contribute.  The possibility that long string excitations exist even
below wavelength $\Gamma G\mu t$ is discussed in Appendix
\ref{sec:appendix-tiny}.

The numerator of Eq.~(\ref{eq:nr}) has little dependence on $\alpha$
except for the largest loops.  We can integrate the results shown in
Fig.~\ref{fig:a3-2frad} to get
\be\label{eq:nrnum}
\int_0^\infty{\alpha'}^{3/2} \fr(\alpha')d \alpha' \approx1.03,
\ee
which is the normalization for all loops with $\alpha$ significantly
below 0.05 .  Note that this result is quite different from what one
would get by normalizing $n(\alpha)$ using the flow of energy from
long strings.  The scaling rate of energy flow into loops in the
radiation era is \cite{BlancoPillado:2011dq} $2(1-\langle
v_\infty^2\rangle)/\gamma_r^2 \approx 53$.  Energy conservation implies
that this is equal to $\int \alpha'\sqrt{p'^2+1}\fr d\alpha'dp'$.  If
one incorrectly takes this to be $\int \alpha' \fr d\alpha'$, and
loops to be created mainly at $\alpha = 0.05$, one mistakenly
concludes $\int {\alpha'}^{3/2}\fr(\alpha')d \alpha' \approx
53\sqrt{0.05} \approx 12$, more than an order of magnitude too large.
This error results from neglecting two important effects: (i) Even
though most loops of a given size at any given time originally formed
with $\alpha \approx 0.05$, most energy leaving the long string
network goes into smaller loops (see Ref.~\cite{Dubath:2007mf}), and (ii) most of the energy leaving
the network goes into loop kinetic energy, which is lost to
redshifting.  The latter effect is less important, since the large
loops which contribute most to $\nr(\alpha)$ are also the slowest.
These effects lead to a change in constraints resulting from
non-detection of gravity waves, which we discuss in
Sec.~\ref{sec:gw}.

Using Eqs.~(\ref{eq:nr}) and (\ref{eq:nrnum}), and a delta-function approximation for
$\fr(\alpha)$ peaked at the typical loop production size $\alpha =
0.05$, we can approximate the loop spectrum by
\beq\label{eq:approxnr}
\nr(\alpha) =  \frac{0.52\, \Theta(0.05-\alpha)}{\left(\alpha + \Gamma G\mu/2\right)^{5/2}},
\eeq
where $\Theta$ is the Heaviside step function.

\begin{figure}
   \centering
   \includegraphics[width=6.0in]{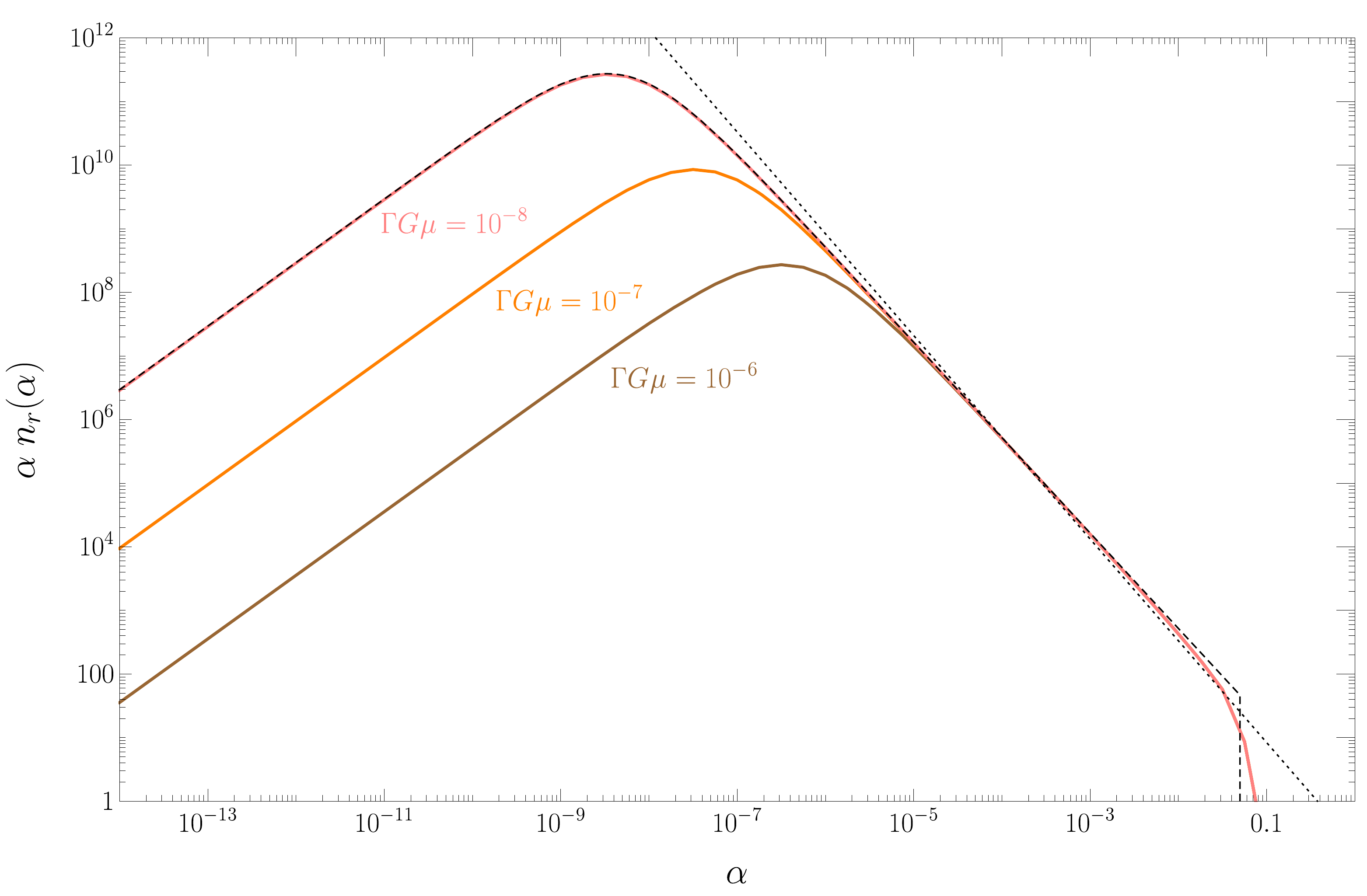} 
   \caption{The loop number density during the radiation era for
     various values of the evaporation rate $\Gamma G\mu$.  Solid
     lines are computed from simulation data for $f$; the dashed line
     is the analytic approximation of Eq.~(\ref{eq:approxnr}) for
     $\Gamma G\mu = 10^{-8}$.  As discussed in the text, the
     non-scaling peak in loop production is invisible here.  The
     dotted line represents the simulation fit in Eq.~(3) of RSB
     \cite{Ringeval:2005kr}, which agrees rather well with the results
     presented here.}
   \label{fig:nrad}
\end{figure}

In Fig.~\ref{fig:nrad}, we plot this $\nr(\alpha)$ and the loop
spectrum computed from simulation data without approximations.
There are several universal features worth pointing out.  For $\alpha
\gtrsim 0.05$, there are no (non-self-intersecting) loops, since none are produced
larger than this size.  At smaller sizes ($\alpha \lesssim 10^{-3}$),
the number density grows with a slope which is universal, given by
$\alpha\, n(\alpha) \approx 0.6 \alpha^{-3/2}$.  The peak number density
per log $\alpha$ occurs at $\alpha = \Gamma G\mu/3$, below which
$\alpha\,n(\alpha)\propto \alpha$.

Using Eq.~(\ref{eq:approxnr}), we can find the total loop number
density in scaling units, 
\be\label{eq:totalr}
\nr = \int \nr(\alpha) d\alpha \approx 0.97 (\Gamma G\mu)^{-3/2},
\ee
the average loop mass,
\ba
\langle\alpha\rangle_r= \frac{\int \alpha\nr(\alpha) d\alpha}{\int \nr(\alpha) d\alpha} \approx \Gamma G\mu,
\ea
and the loop matter density,
\ba
\langle \alpha \rangle \nr \approx 0.97 (\Gamma G\mu)^{-1/2}.
\ea
The energy density in long strings \cite{BlancoPillado:2011dq} is
about 44 in scaling units, so the energy density in loops is larger by
the factor
\be
\frac{\rho_r^{\rm loops}}{\rho_r^{\infty}} \approx 100\sqrt{\left(\frac{50}{\Gamma}\right)\left(\frac{10^{-9}}{G\mu}\right)}.
\ee
Current bounds on the string tension (see Sec.~\ref{sec:gw}) imply the
radiation era loop energy density is nearly two orders of magnitude
larger than the long-string energy density, at a minimum.

\begin{figure}
   \centering
   \includegraphics[width=4.0in]{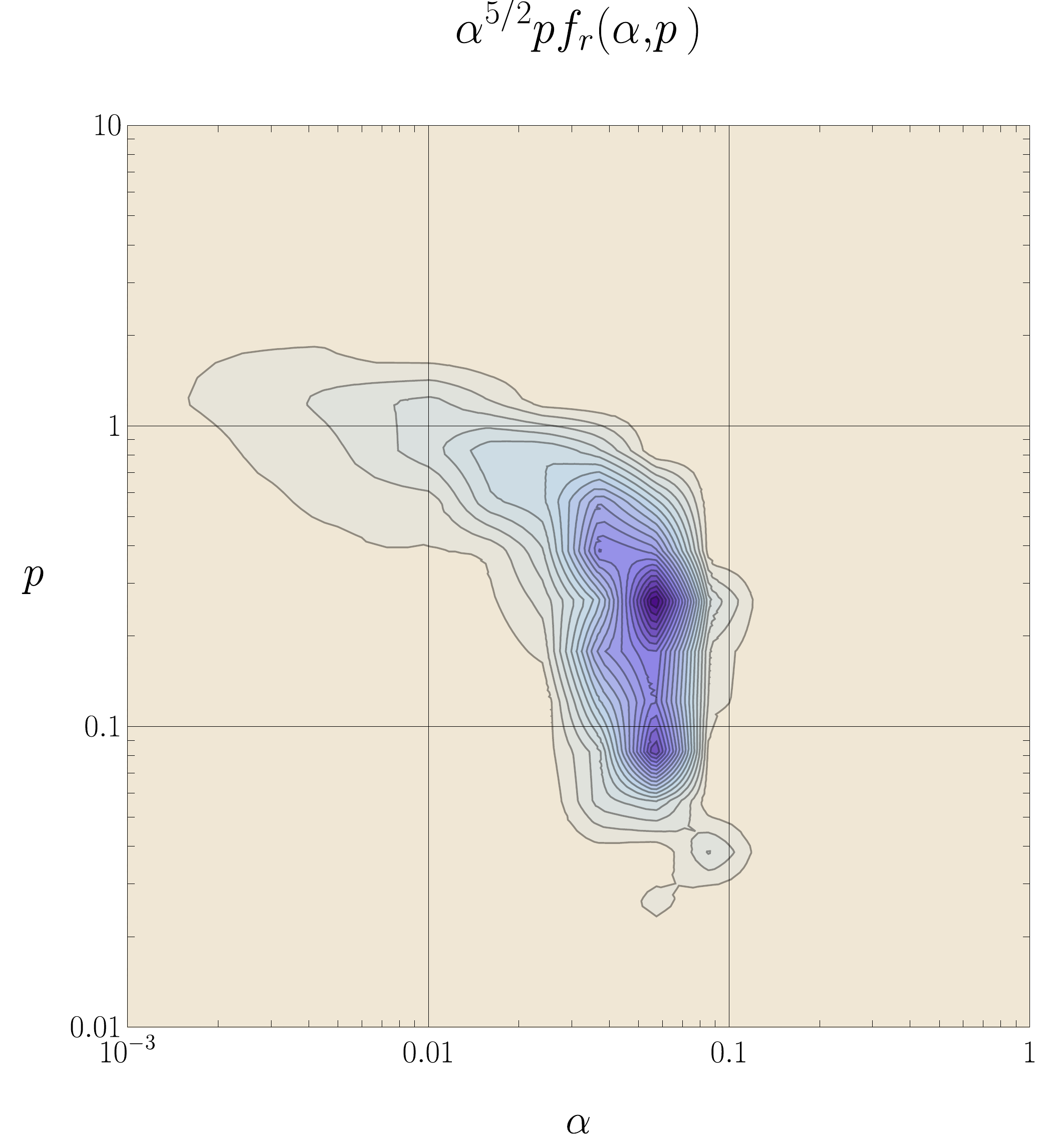}
   \caption{Linearly spaced contours of $\alpha^{5/2}p\fr(\alpha,p)$,
     which corresponds to the quantity integrated in Eq.~(\ref{eq:nvr2}).
     Expansion of the universe causes any given point to flow downward and to the 
    left at a slope of $1/2$.  }
   \label{fig:a3-2fcontourrad}
\end{figure}

Now we consider the momenta of the loops.  Returning to
Eq.~(\ref{eq:nv}), for the radiation era we find
\be\label{eq:nvr}
\nr(\alpha,p) = \frac{\int_\alpha^\infty (\alpha' + \Gamma G\mu/2)^2
\fr(\alpha',P')d\alpha'}{2(\alpha + \Gamma G\mu/2)^3},
\ee
with $t' = t(\alpha + \Gamma G\mu/2)/(\alpha' + \Gamma G\mu/2)$ and
thus
\be\label{eq:pm}
P' = \left(\frac{\alpha' + \Gamma G\mu/2}{\alpha + \Gamma G\mu/2}\right)^{1/2} p.
\ee
By writing Eq.~(\ref{eq:nvr}) as
\be\label{eq:nvr2}
 \nr(\alpha,p) = \frac{\int_\alpha^\infty (\alpha' + \Gamma G\mu/2)^{3/2}
 \fr(\alpha',P')\frac{\partial P'}{\partial p} d\alpha'}{2(\alpha + \Gamma G\mu/2)^{5/2}},
\ee
we can see that integration over $dp$ returns Eq.~(\ref{eq:nr}).
The quantity being integrated is represented in Fig.~\ref{fig:a3-2fcontourrad}.

Even though the vast majority of loops are very small ones emitted
with ultra-relativistic speeds, they are suppressed by the
$\alpha'^{3/2}$ factor, so their contribution to the numerator of
Eq.~(\ref{eq:nvr}) is small.  Some loops do exist with $p\sim 2$ and
$\alpha\sim 10^{-3}$.  For these relativistic loops, the mass loss
formula of Eq.~(\ref{eq:nrm}) is not accurate.  However, within a few
Hubble times, before they have lost a significant fraction of their
mass, these loops will be redshifted to nonrelativistic speeds.  Thus
the use of the nonrelativistic approximation in Eq.~(\ref{eq:nrm}) is
justified.

\begin{figure}
   \centering
   \includegraphics[width=6.0in]{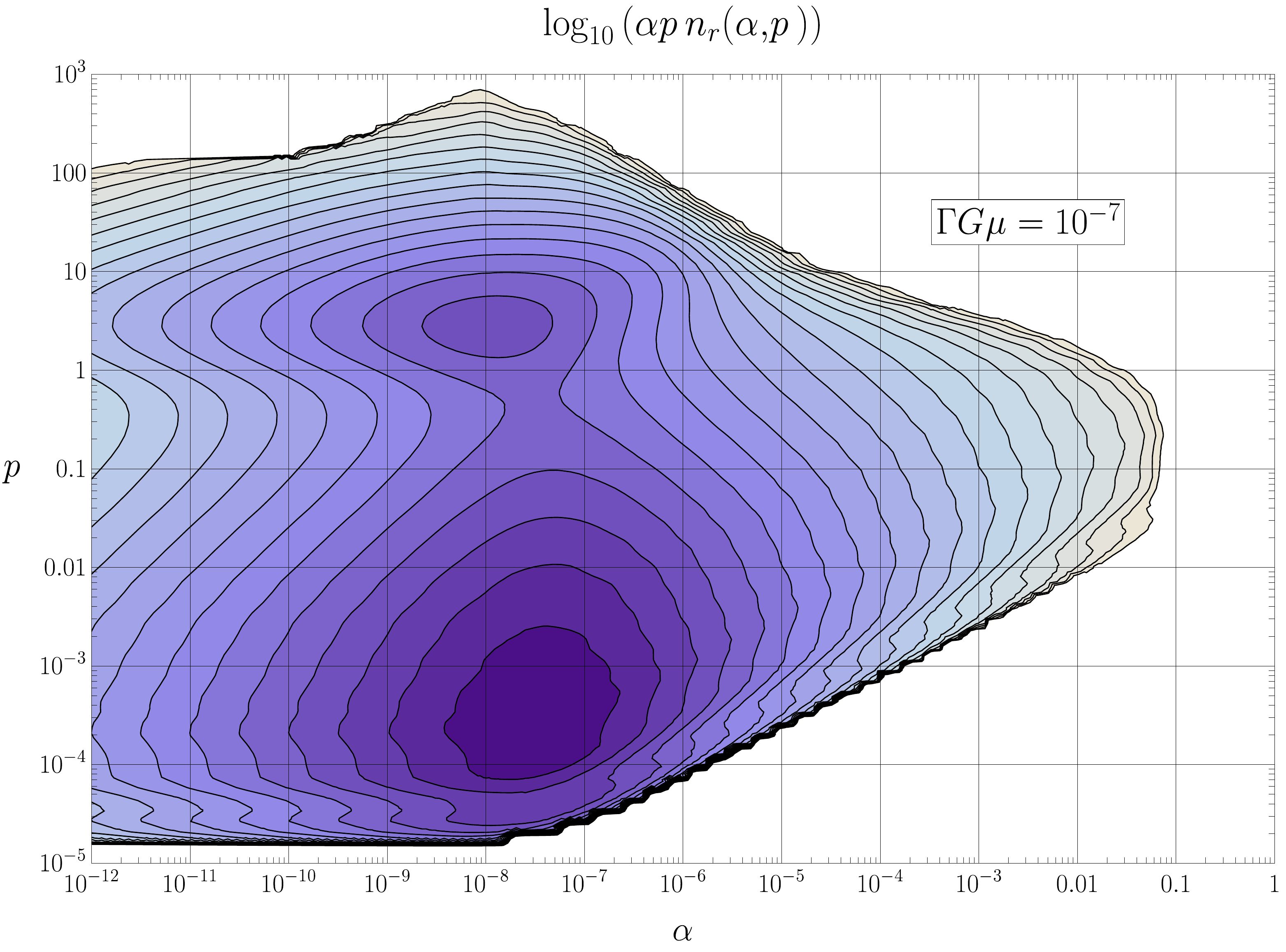} 
   \caption{The mass and momentum spectrum of loops in the radiation
     era, using simulation data for the loop production function and
     the exact flow.  Contours are of $\log_{10}(\alpha p
     \nr(\alpha,p))$, with each contour representing one-half of an
     orders of magnitude, ranging from $10^0$--$10^9$.  Notice that the
     non-scaling peak at the top of the figure is subdominant by a few
     orders of magnitude; The vast majority of loops are in the
     scaling peak at $\alpha \approx \Gamma G\mu/3$, $p \approx
     \sqrt{\Gamma G\mu}$.}
   \label{fig:ncontourrad}
\end{figure}

The number density computed using simulation data for the loop production function, and the
evaporation rate $\Gamma G\mu = 10^{-7}$ is shown in Fig.~\ref{fig:ncontourrad}. 

\begin{figure}
   \centering
   \includegraphics[width=4.0in]{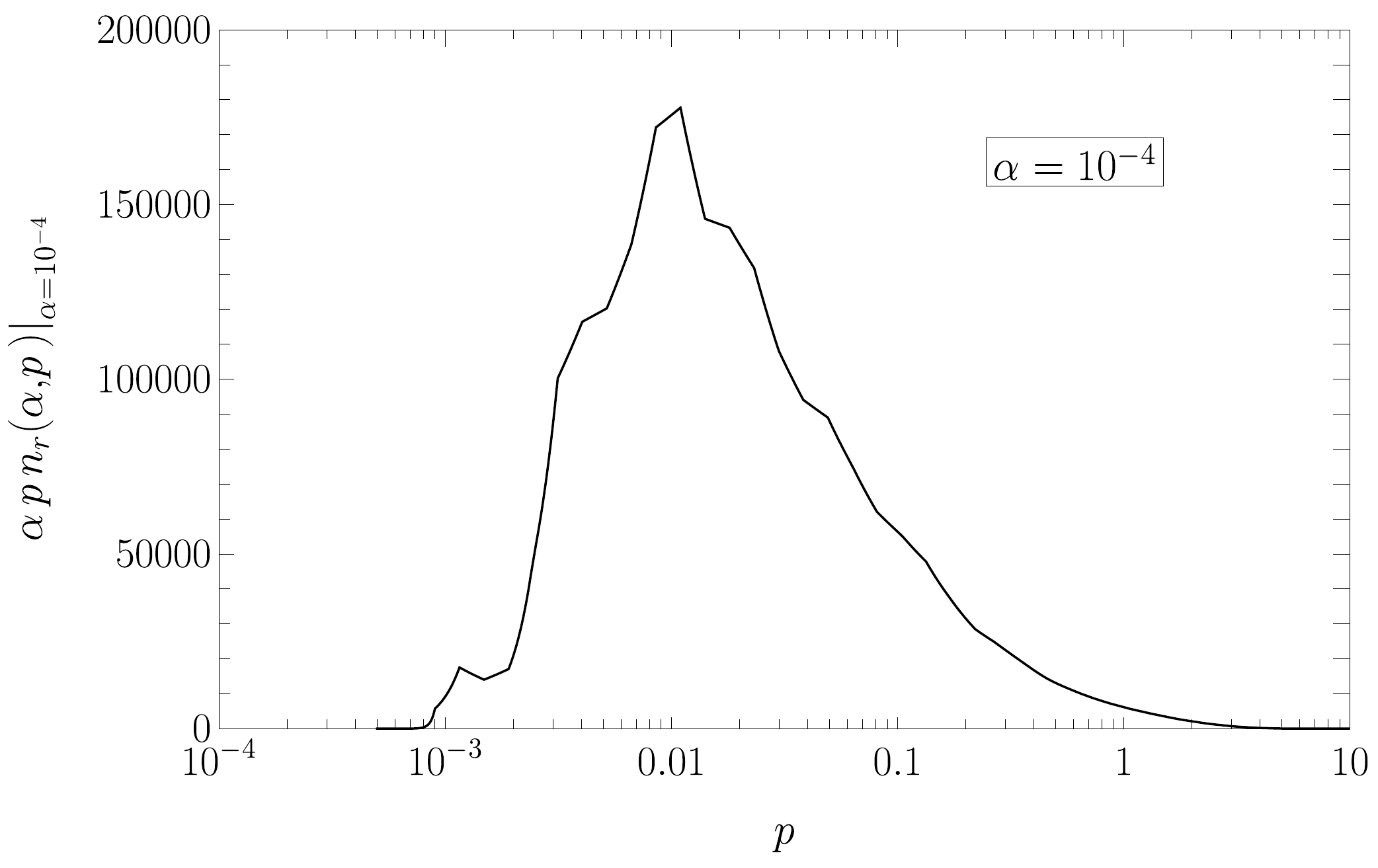} 
   \caption{The momentum distribution of loops for a slice with
     constant $\alpha = 10^{-4}\gg \Gamma G\mu$ during the radiation
     era.  For a different $\alpha \gg \Gamma G \mu$, the horizontal
     position of features in the graph is proportional to
     $\alpha^{1/2}$ and the vertical position proportional to
     $\alpha^{-3/2}$.}
   \label{fig:nvrad}
\end{figure}

Now consider the momentum distribution of loops with a fixed $\alpha$.
This is shown in Fig.~\ref{fig:nvrad} for $\alpha = 10^{-4} \gg \Gamma
G \mu$.
If $\alpha \lesssim 10^{-3}$, we can approximate the lower limit of integration as $\alpha=0$
in the numerator of Eq.~(\ref{eq:nvr}), which thus depends on $\alpha$
only through $p'$.  Then the distribution for a different $\alpha$ can
be found by
\be
\nr(\alpha_2,p) = \left(\frac{\alpha_1 + \Gamma G\mu/2}{\alpha_2 +
  \Gamma G\mu/2}\right)^{3} \nr\bigg(\alpha_1,p\sqrt{\frac{\alpha_1 +
    \Gamma G\mu/2}{\alpha_2 + \Gamma G\mu/2}}\bigg),
\ee
so the shape of the distribution of $\nr(\alpha,p)$ at fixed $\alpha \lesssim 10^{-3}$
is shifted and scaled but not changed in shape under a change in
$\alpha$.

The majority of loops (by either count or energy) at any given time
have sizes $\alpha\lesssim \Gamma G\mu$.  For such $\alpha$, the peak in
Fig.~\ref{fig:nvrad} is shifted to $p\sim\sqrt{\Gamma G\mu}$, so this
is the typical loop speed.

\subsection{Matter era}
Here we consider loops during the matter era.  There are always some
loops produced during the matter era, but initially they are dwarfed by relic loops from the radiation era.  The exception
is loops larger than $\alpha \approx 0.05 d_{h}(\teq)/d_h(t)$, of which none are from the radiation era.

The largest loops produced in the radiation era have size about
$0.1\teq$ and thus survive until time about $0.1\teq/(\Gamma G\mu)$.
Thus loops remain today from the radiation era if 
\be\label{eq:radiationcriterion}
\Gamma G\mu \lesssim \frac{0.1 \teq}{t_0}  \approx 3.6\times 10^{-7}.
\ee

When we use scaling units in the matter era we will always scale by
the appropriate power of $d_h \equiv 3t$, which would be the horizon
distance for a universe which had always been matter dominated, rather
than taking into account the previous radiation era.

In the matter era, Eq.~(\ref{eq:n}) becomes
\be\label{eq:nm}
\nm(t,\alpha) = \frac{\int_\alpha^{\alphaeq} \left(\alpha' + \Gamma G\mu\right/3)\fm(\alpha')d\alpha'}
{3\left(\alpha + \Gamma G\mu\right/3)^{2}}.
\ee
For a universe which is always been matter dominated, $\Alphaeq =\infty$,
but in the real situation in there is a cutoff, because $\alpha'$
should not be so large that the corresponding $t'<\teq$, giving
\be\label{eq:alphamax}
\Alphaeq\approx \frac{t}{\teq} (\alpha +\Gamma G\mu/3).
\ee
For the most part this cutoff is unimportant, since loops surviving from the radiation era will overwhelm the
subpopulation affected by finite $\Alphaeq$. 

\begin{figure}
   \centering
   \includegraphics[width=5.0in]{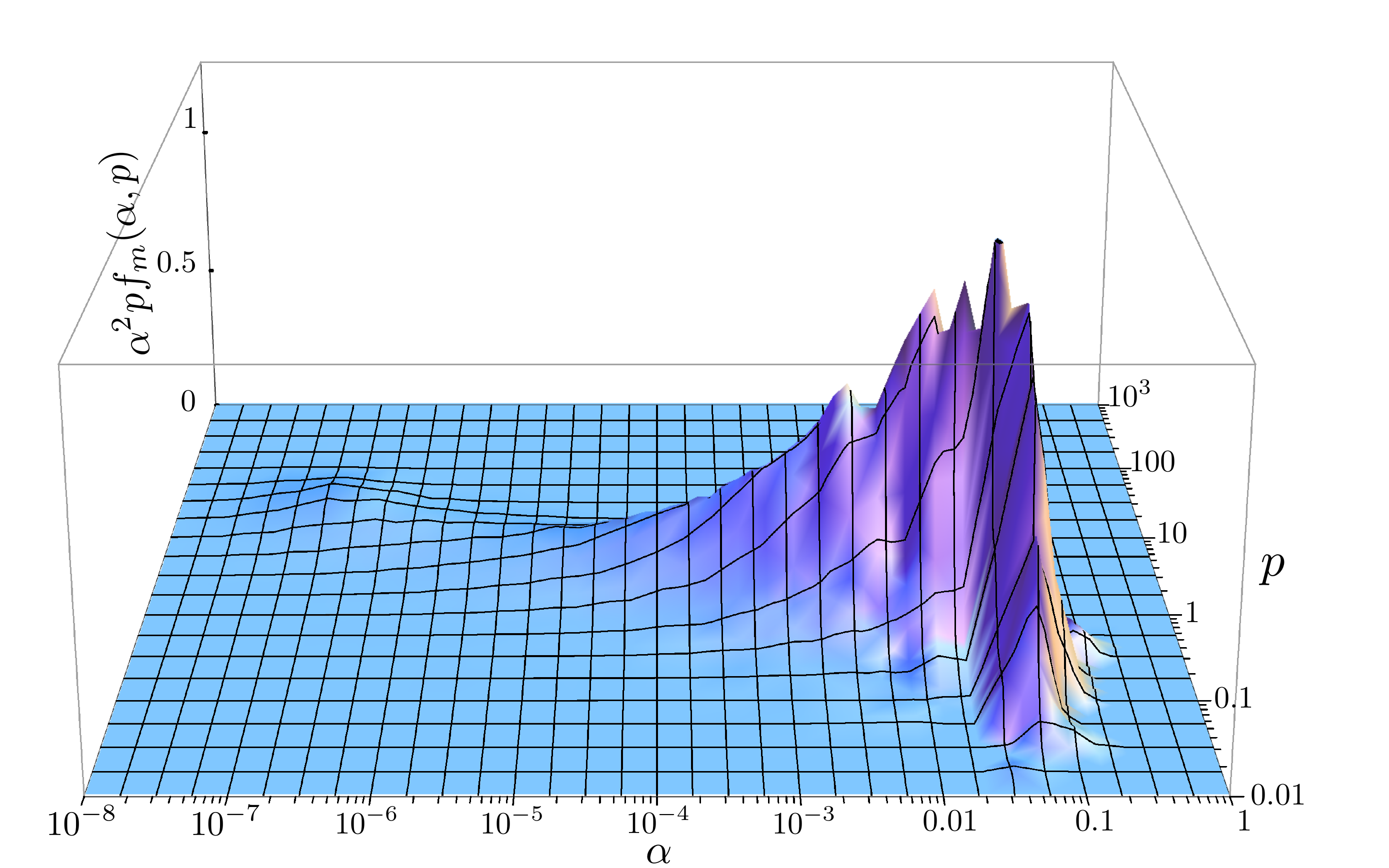} 
   \caption{The mass production function during the matter era.  The
     non-scaling peak is almost invisible on a linear scale.}
   \label{fig:afmat2d}
\end{figure}

\begin{figure}
   \centering
    \hspace{-.2in}
   \subfigure{
       \includegraphics[width=3.12in]{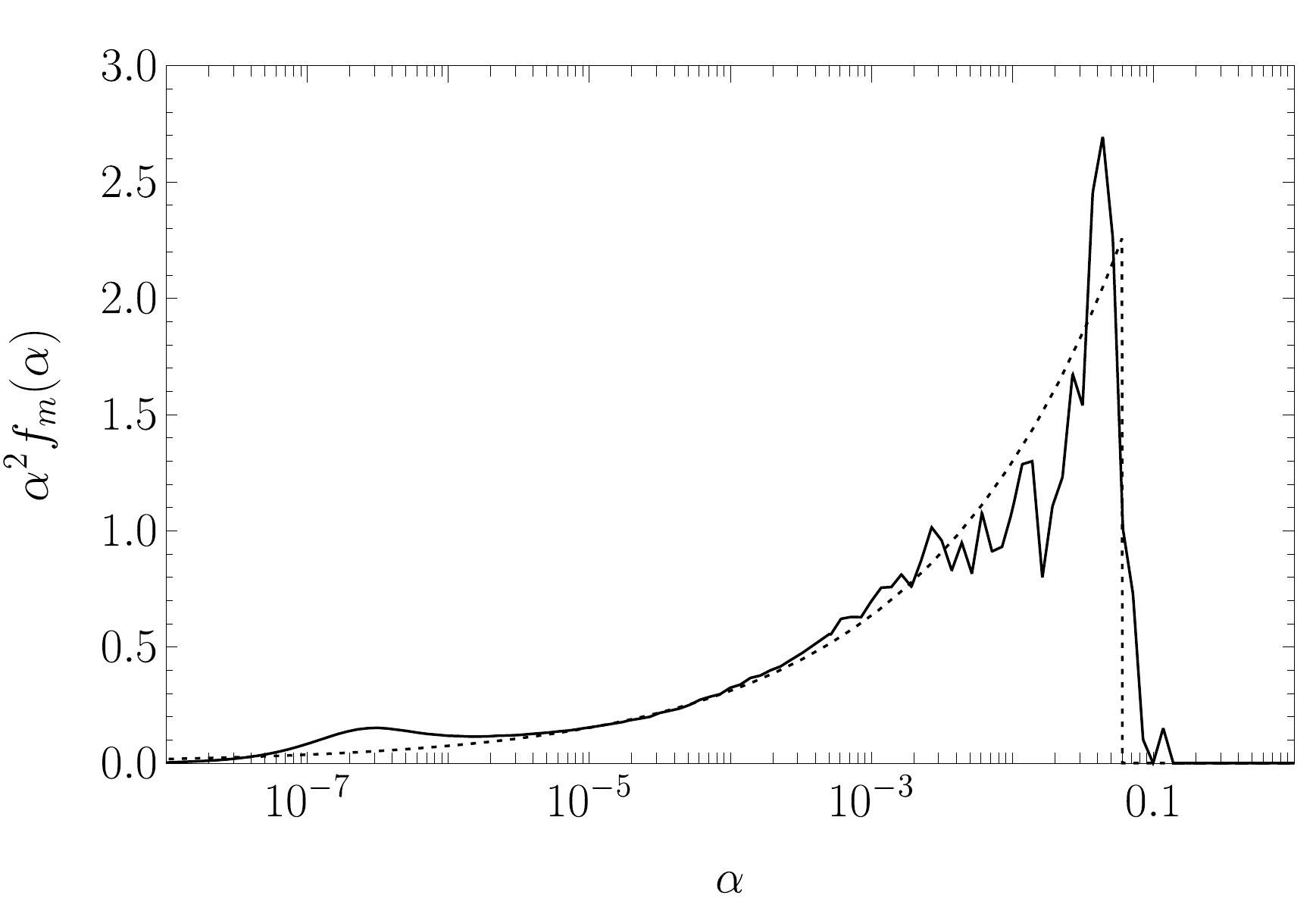} 
     }
        \subfigure{
   \includegraphics[width=3.18in]{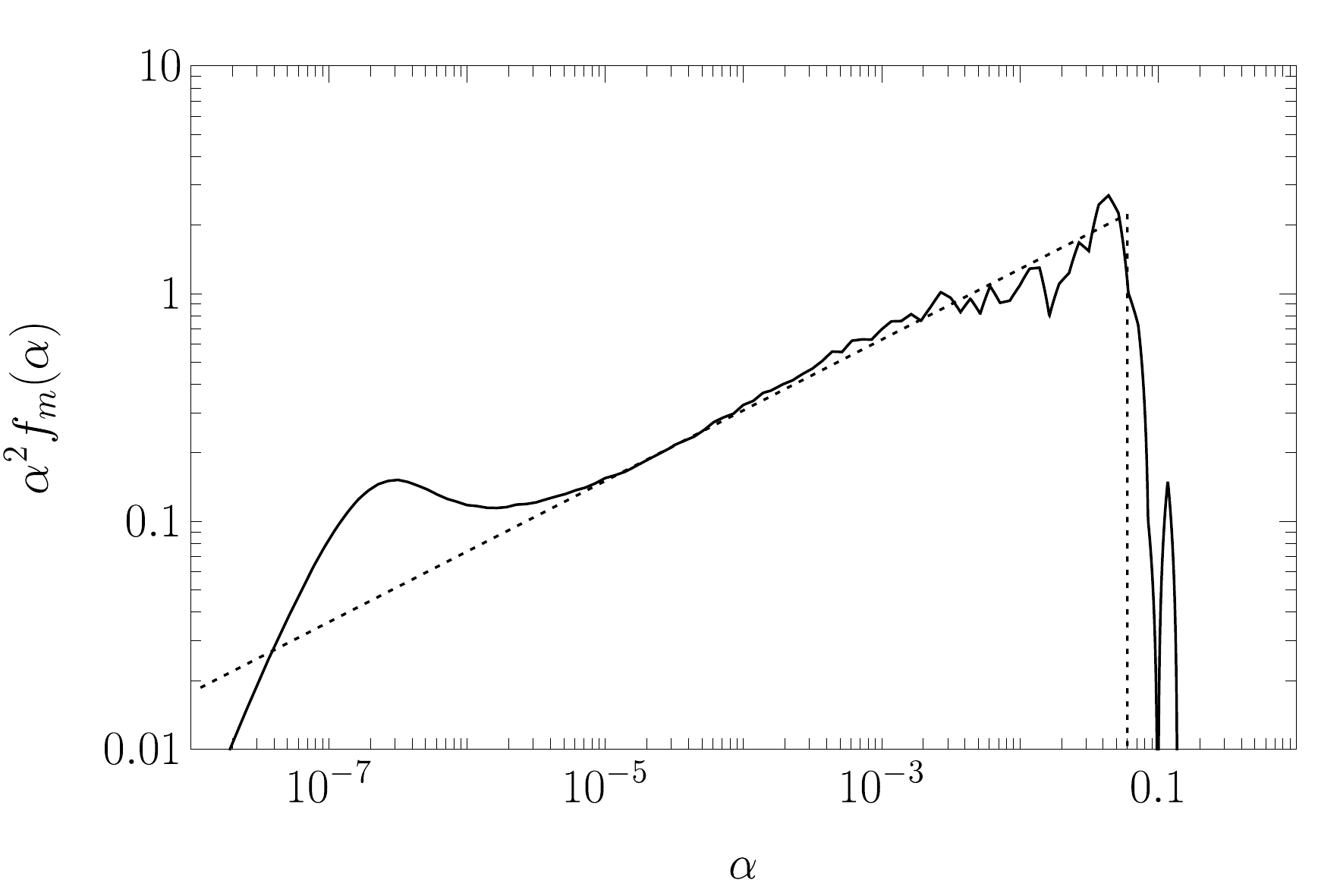} 
     }
   \caption{The loop production rate $f(\alpha)$ scaled by $\alpha^2$.
     The non-scaling peak represents a subdominant contribution to the
     total number of loops, although it may contribute significantly
     to the small loop sub-population.  The dotted line shows the
     approximation of Eq.~(\ref{eq:mfit}). Note that the non-scaling peak would be more
     prominent in these graphs if we were to use the scaling energy instead of the rest 
     mass of the loops.  See the corresponding figures in \cite{BlancoPillado:2011dq}. This is
     due to the fact that most of the energy of the small loop population is in kinetic energy, not rest mass.}
   \label{fig:afmat1d}
\end{figure}

The numerator in Eq.~(\ref{eq:nm}) is now just the total mass
production function; there is no additional half power of $\alpha'$ as
there was during the radiation era.  In Figs.~\ref{fig:afmat2d} and
\ref{fig:afmat1d}, we show this quantity from 3 simulation runs of
size 1000 starting at conformal time\footnote{In previous work
  \cite{BlancoPillado:2011dq}, we used starting time 4.5 in the matter
  era.  The present choice of 9.0 makes the large-loop part of the
  loop production function closer to its scaling value at early
  times.} 9 and running for time 500 for a dynamic range of 56.
\cite{BlancoPillado:2011dq}.

The simulations have reached the scaling regime, in the sense that the
non-scaling peak at $\alpha \sim 10^{-7}$ is subdominant.  The behavior of tiny loops is again irrelevant for the
total number of loops, thanks entirely to the large kinetic energies of small loops.
As in the radiation era, if
gravitational damping smooths long strings below the scale $\Gamma G\mu t$,
then much smaller $\alpha'$ make no contribution.
The possibility of smaller smoothing scales is discussed in Appendix
\ref{sec:appendix-tiny}.

In Eq.~(\ref{eq:nm}), the smaller power of $\alpha'$ multiplying the loop production function 
means that loop production effectively happens over a broader hierarchy of scales, as compared with the radiation era.  Hence, the number
density $n(\alpha)$, which is the integral of this broad production, will approach the universal power law $\alpha^{-1}$ much more slowly, having
a steeper slope at larger $\alpha$.
This may explain why the numerical fit of RSB is steeper than the universal slope, since the fit was performed at large $\alpha$.

\begin{figure}
   \centering
   \includegraphics[width=5.5in]{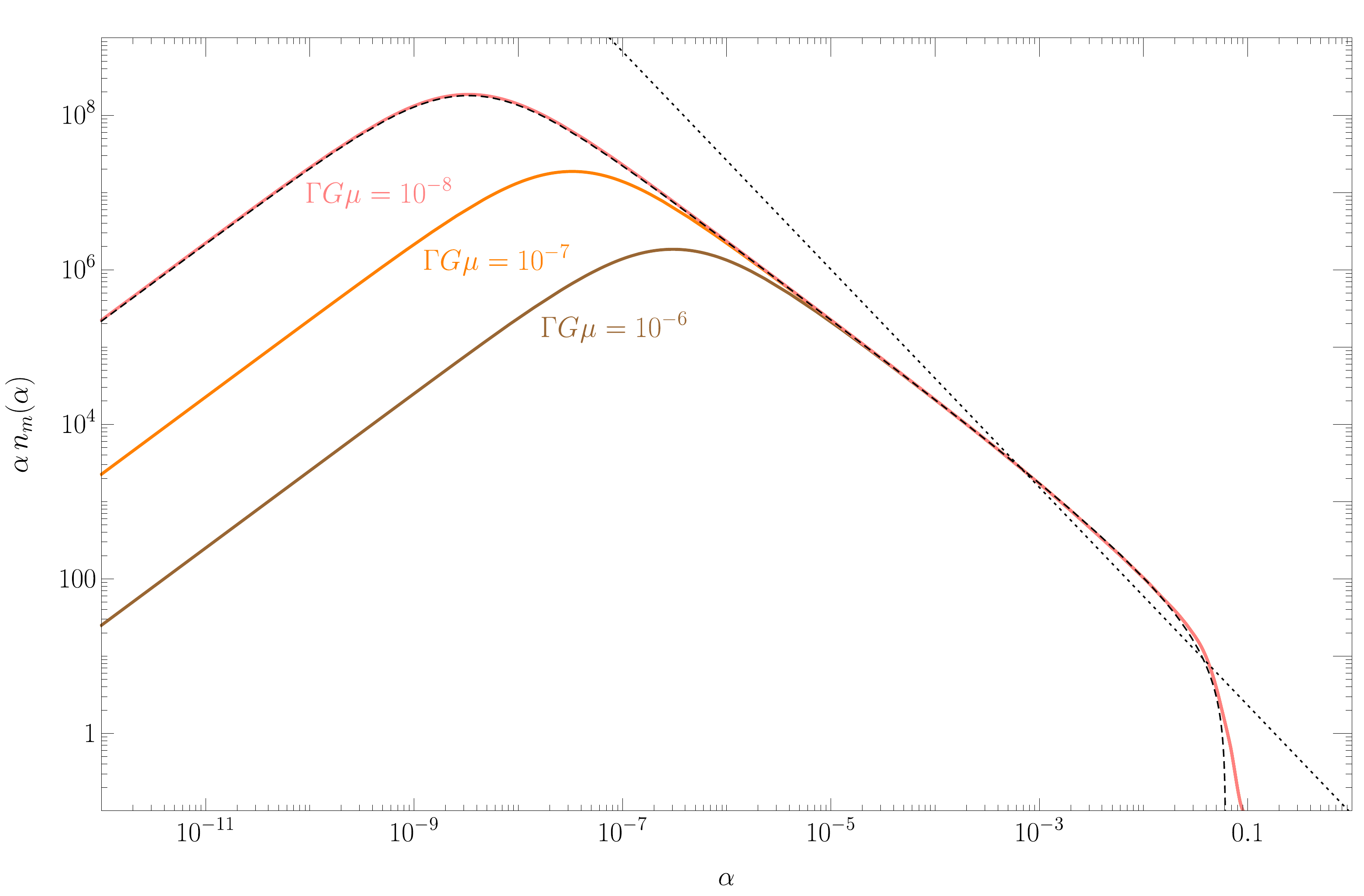} 
   \caption{The loop number density during the matter era for various
     values of the evaporation rate $\Gamma G\mu$.  Solid lines use
     pure simulation data for the loop production function and the
     non-relativistic loop evaporation rate.  The dashed line is the
     analytic approximation of Eq.~(\ref{eq:approxnmat}) for $\Gamma
     G\mu = 10^{-8}$.  The dotted line represents the fit (roughly for
     $10^{-3}\lesssim \alpha \lesssim 10^{-1}$) in Eq.~(3) of RSB
     \cite{Ringeval:2005kr}.  Although this fit  cannot be extrapolated to
     small $\alpha$, the data appear consistent with our results.}
   \label{fig:nmat}
\end{figure}

An analytic approximation for the matter era loop production function
is
\beq\label{eq:mfit}
\fm(\alpha) \approx \frac{5.34}{\alpha^{1.69}}\Theta(0.06-\alpha)
\eeq
shown in Fig.~\ref{fig:afmat1d}.  Neglecting any possible contribution
from $\alpha<\Gamma G \mu$,  Eq.~(\ref{eq:mfit}) results in the analytic
approximation for the scaling loop spectrum,
\beq\label{eq:approxnmat}
\nm(\alpha) \approx \frac{2.4 - 5.7\alpha^{0.31}}{\left(\alpha + \Gamma G\mu/3\right)^2},
\eeq
and its integral
\be\label{eq:totalm}
\nm \approx 7.2 (\Gamma G\mu)^{-1}.
\ee
We illustrate $\nm(\alpha)$ in Fig.~\ref{fig:nmat}, using simulation
data and the non-relativistic loop evaporation rate.
Including the transient effect of $\Alphaeq$ gives
\beq\label{eq:transientnmat}
\nm(t,\alpha) \approx \frac{\left(2.4 - 5.7\alpha^{0.31}\right)\Theta(0.06 - \alpha )-\left(2.4 - 5.7\Alphaeq^{0.31}\right)\Theta(0.06 - \Alphaeq )}{\left(\alpha + \Gamma G\mu/3\right)^2}.
\eeq

\begin{figure}
   \centering
   \includegraphics[width=5.0in]{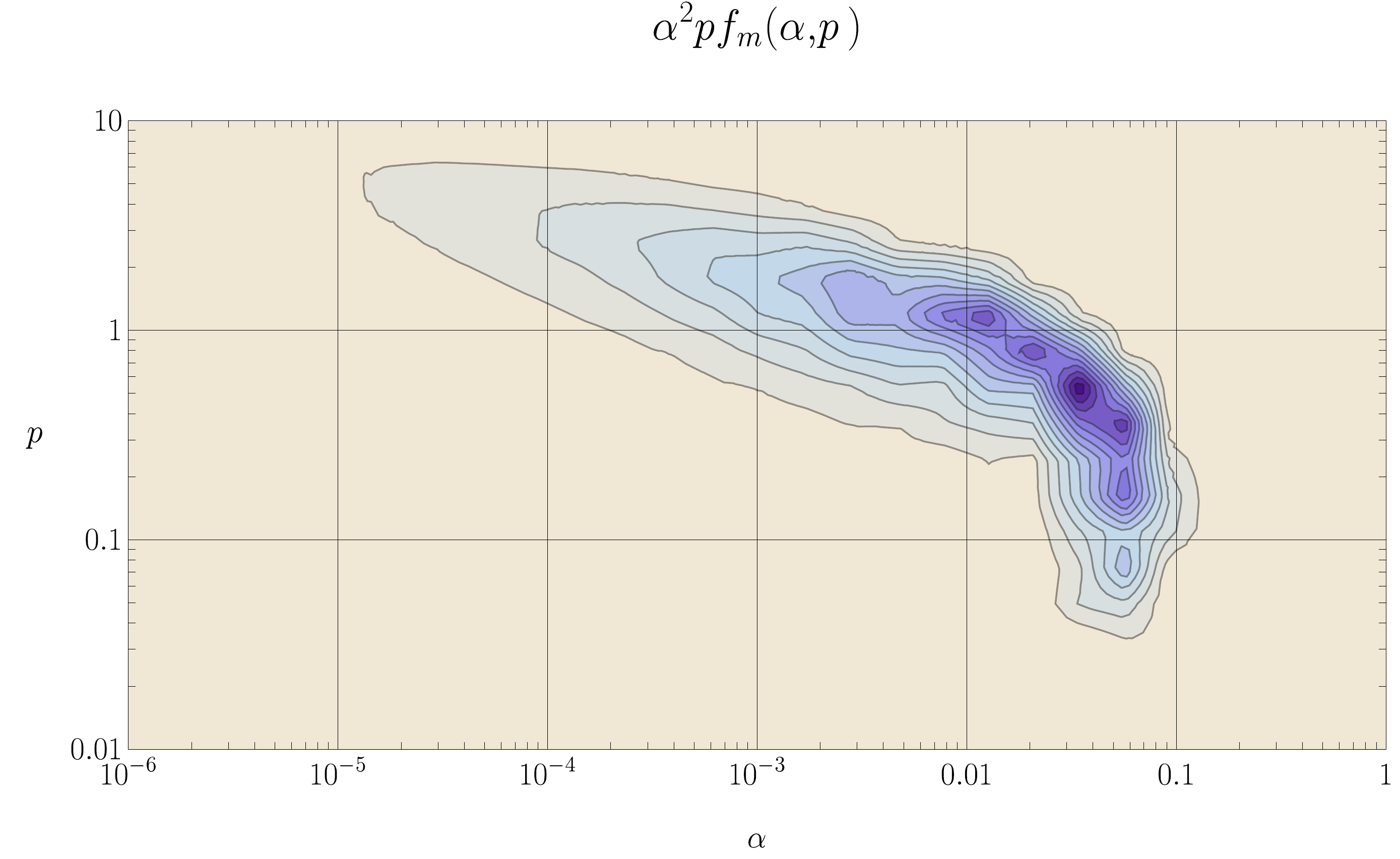}
   \caption{A contour plot of the mass production function
    during the matter era.  This is the quantity relevant for determining
    the matter era loop number density.  Expansion of
     the universe causes any given point to flow downward and to the
     left at a slope of $2/3$. Contours are of $\alpha^2 p f_m(\alpha,p)$, spaced linearly from $0.1$--$1.3$.}
   \label{fig:afcontourmat}
\end{figure}

\begin{figure}
   \centering
   \includegraphics[width=6.0in]{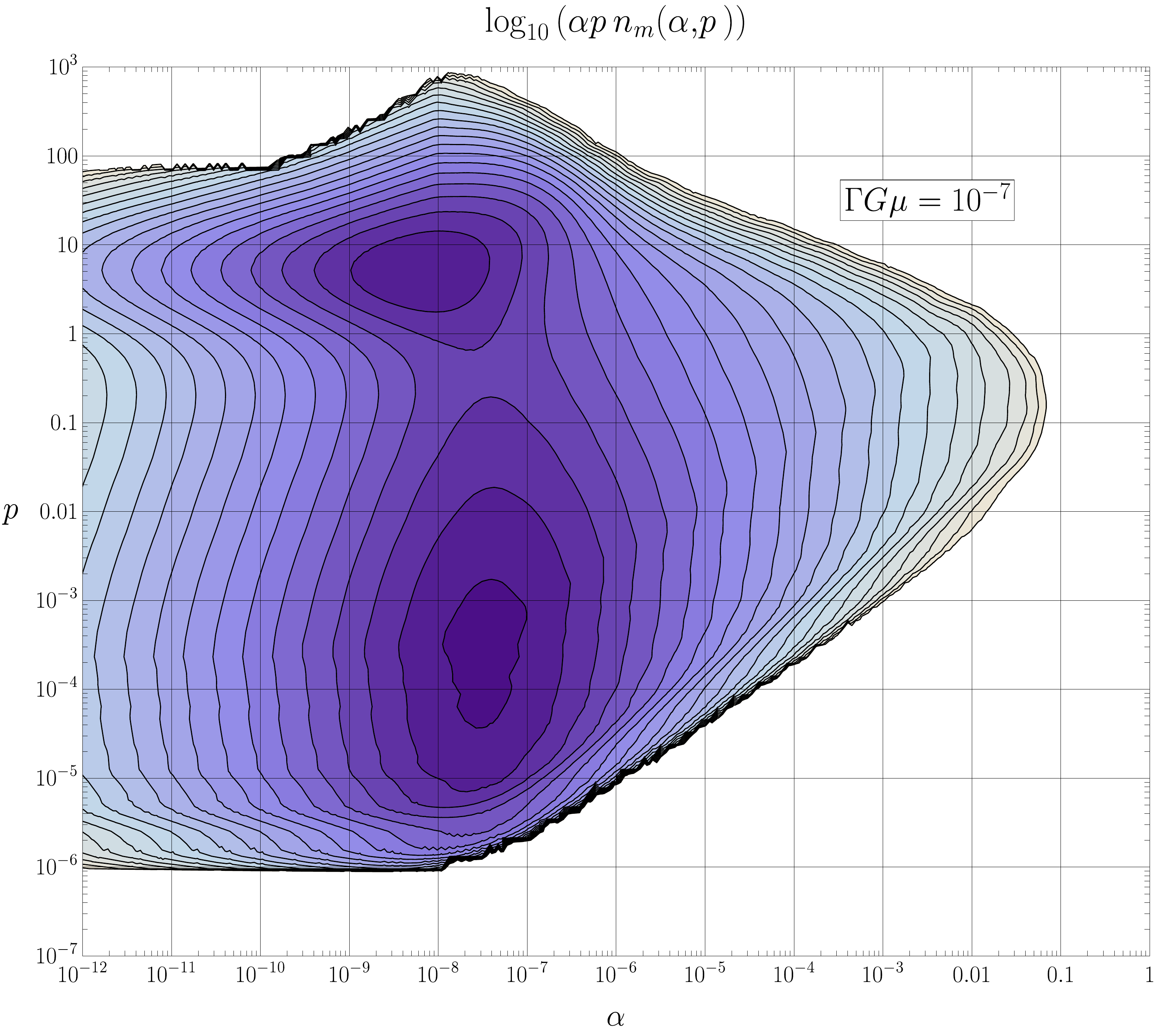} 
   \caption{The mass and momentum spectrum of loops in a matter dominated universe, using simulation data for the loop production function and the exact flow.  Contours are of $\log_{10}(\alpha p \nr(\alpha,p))$, with
   each contour representing one-third of an order of magnitude, ranging from $10^0$--$10^{19/3}$.  Notice that the non-scaling peak at the top of the figure
   is subdominant by an order of magnitude;  The majority of loops are in the scaling peak at $\alpha \approx \Gamma G\mu/3, p \approx 10\left(\Gamma G\mu\right)^{2/3}$.}
   \label{fig:ncontourmat}
\end{figure}

Now we consider momenta in the matter era.  Equation~(\ref{eq:nv}) gives
\be\label{eq:nvm}
\nm(t,\alpha,p) = \frac{\int_\alpha^{\alphaeq} (\alpha' + \Gamma G\mu/3)^{5/3}
\fr(\alpha',P')d \alpha'}{3(\alpha + \Gamma G\mu/3)^{8/3}},
\ee
with $t' = t(\alpha + \Gamma G\mu/3)/(\alpha' + \Gamma G\mu/3)$ and thus
\be\label{eq:pr}
P' = \left(\frac{\alpha' + \Gamma G\mu/3}{\alpha + \Gamma G\mu/3}\right)^{2/3} p.
\ee
By writing this as
\be\label{eq:nvm2}
\nm(t,\alpha,p) = \frac{\int_\alpha^{\alphaeq} (\alpha' + \Gamma G\mu/3)
\fr(\alpha',P')\frac{\partial P'}{\partial p}\,d\alpha'}{3(\alpha + \Gamma G\mu/3)^2},
\ee
we can see that integration over $dp$ returns Eq.~(\ref{eq:nm}).  The object being integrated is represented in Fig.~\ref{fig:afcontourmat}.
In Fig.~\ref{fig:ncontourmat}, we show the distribution of loops in a
matter-dominated universe, including relativistic effects on loop evaporation.

In this case, there is a substantial tail extending to high boost.
When we use the nonrelativistic formula for mass lost to gravity
waves, Eq.~(\ref{eq:nrm}), we overestimate the change by a factor
$\gamma$.  This overestimate occurs until the loop momentum per unit
mass falls from $p$ to of order unity.  If the loop is created at time $t_i$,
this happens at $\tnr$ where $a(\tnr)/a(t_i) = p$, or $\tnr = p^{3/2}
t_i$.  For large $p$, the overestimate of the total emitted mass is
$\Gamma G \mu^2 \tnr = \Gamma G \mu^2 p^{3/2} t_i$.  This is of little
consequence providing it is much less than the the original loop mass
$\alpha d_h \mu$ = $3 \alpha t_i \mu$.  Thus we consider what fraction
of the integral in numerator of Eq.~(\ref{eq:nm}) is made up of
high-momentum loops with
\be
\Gamma G \mu p^{3/2} > 3 \alpha.
\ee
Even for very heavy strings with $\Gamma G \mu = 10^{-5}$ this
fraction only is about 11\%, and for $\Gamma G \mu = 10^{-7}$ it is
about 5\%, so there is some error from this effect, but not a very
large one.

\begin{figure}
   \centering
   \includegraphics[width=4.0in]{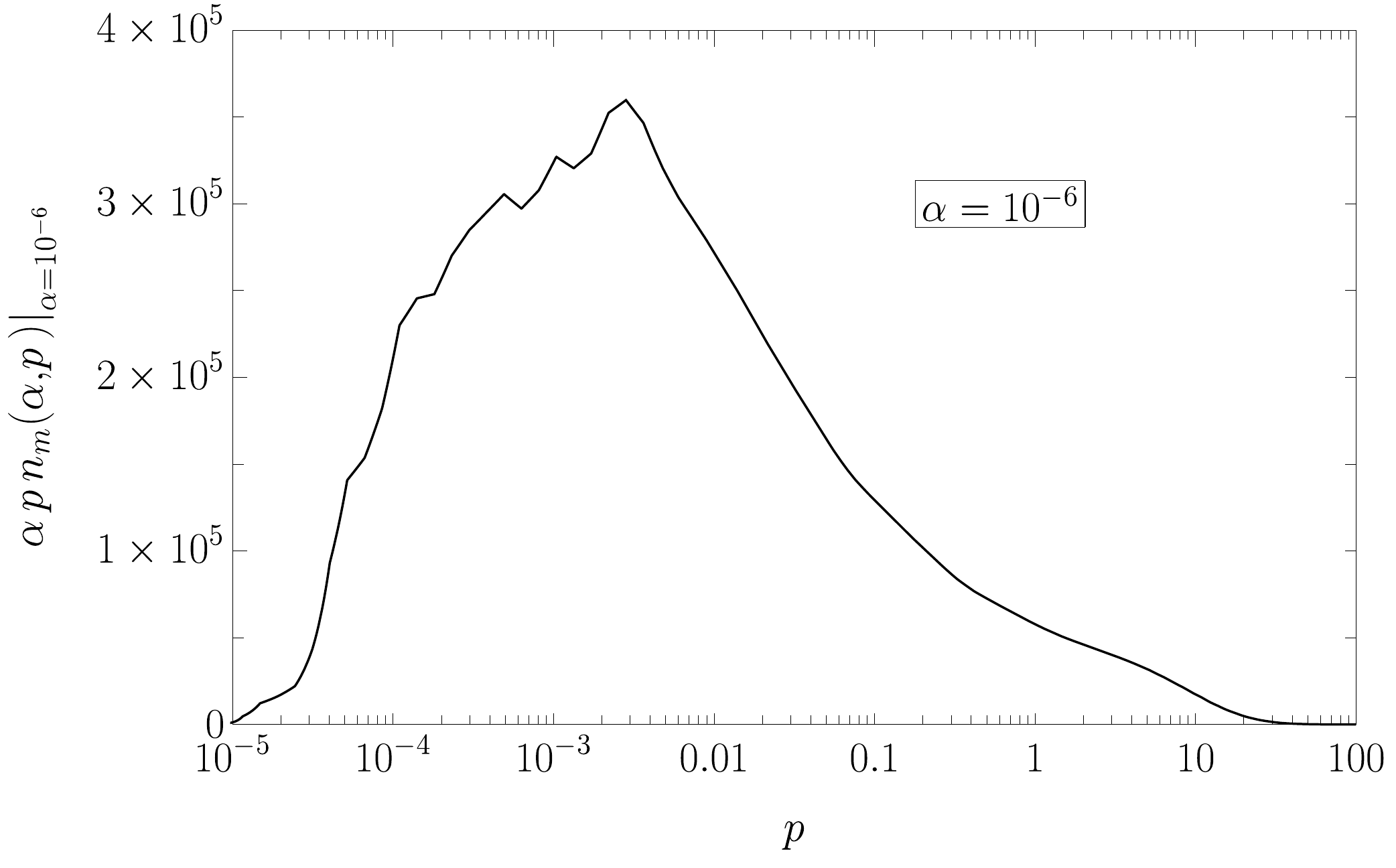} 
   \caption{The momentum distribution of loops with $\alpha =
     10^{-6}\gg \Gamma G\mu$ in a matter dominated universe for a
     slice at constant $\alpha$.  For a different $\alpha \gg \Gamma G
     \mu$, the horizontal position of features in the graph is
     proportional to $\alpha^{2/3}$ and the vertical position
     proportional to $\alpha^{-2}$.}
   \label{fig:nvmat}
\end{figure}

The distribution of velocities for loops with a fixed $\alpha =
10^{-6}$, assuming this is still greater than $\Gamma G \mu$, is shown
in Fig.~\ref{fig:nvmat}.
Note that this is about the smallest $\alpha$ which today describes
loops formed in the matter era.  Smaller loops would be relics from
the radiation era, which we discuss in the next subsection.

The velocity of a typical loop is around $10(\Gamma G\mu)^{2/3}$, but
the distribution is quite broad.

\subsection{Loops surviving from the radiation era}

Most loops in existence during the matter era were produced during the
radiation era.  A loop of size $\alpha$ at time $t$ in the matter era
has mass $m = 3\mu \alpha t$ and thus had mass $\meq = 3\mu
\alpha t + \Gamma G\mu^2 (t-\teq)$ at $\teq$.  Thus, at the end of the
radiation era with $d_h = 2\teq$, this loop had
\be\label{eq:alphaeqmatter}
\Alphaeq = \frac{1}{2}\left[(3 \alpha + \Gamma
  G\mu)\frac{t}{\teq}-\Gamma G\mu\right]
\approx \frac{(3 \alpha + \Gamma G\mu)t}{2\teq}.
\ee
In range of sizes $d\alpha$ at $\teq$ there were $n_r(\alpha)d\alpha$
loops in a horizon volume, $(2\teq)^3$, and thus per comoving volume there
were $n_r(\alpha) d\alpha (\aeq^3/ (2\teq)^3)$.  These loops become
the loops in the same comoving volume at a later time $t$, so
\be
\nr(t>\teq,\alpha) =
\nr(\Alphaeq)\frac{\aeq^3}{a^3(t)}\frac{(3t)^3}{(2\teq)^3}\frac{\partial \Alphaeq}{\partial \alpha}
= \frac{81t^2}{16 \teq^2}\nr(\Alphaeq),
\ee
with $\Alphaeq$ given by Eq.~(\ref{eq:alphaeqmatter}).  This is
time-dependent, since it is not a scaling population.  It should be
pointed out that the apparent discontinuity in $\nr(t,\alpha)$ at
$\teq$ is only due to the discontinuity in our choice of the
``horizon distance'' $\dh = 2t \to 3t$ at $t = \teq$.  The loop number
density is, of course, continuous at $t = \teq$.

Plugging in the analytic approximation from Eq.~(\ref{eq:approxnr}),
we find
\be\label{eq:nrmat}
\nr(t>\teq,\alpha) &\approx& 2.6\frac{t^2}{\teq^2}
\frac{\Theta(0.05-\Alphaeq)}{(\Alphaeq+\Gamma G \mu/2)^{5/2}}\\
&\approx&
0.94\left(\frac{\teq}{t}\right)^{1/2}
\frac{\Theta(0.03(\teq/t)-\alpha-\Gamma G \mu/3)}{(\alpha+\Gamma G \mu/3)^{5/2}}.
\ee

The most important loops are those with $\alpha\sim\Gamma G \mu$.
Ignoring numerical factors, if $\Gamma G \mu \gtrsim \teq/t$ then
these loops were formed in the matter era, and if $\Gamma G \mu
\lesssim \teq/t$ they were formed in the radiation era.  The latter is the case for all realistic values of string tension.
Equation~(\ref{eq:nrmat}) has an extra half power of $\alpha+\Gamma G
\mu/3$ in the denominator as compared to Eq.~(\ref{eq:approxnmat}).
For $\alpha \sim \Gamma G \mu \sim \teq/t$ this is canceled by the
prefactor, but for smaller $\Gamma G \mu$ the number of loops is
enhanced by $\sqrt{\teq/(\Gamma G \mu t)}$ over what it would be in a
purely matter universe.

\section{Stochastic gravitational waves and a bound on $G\mu$}\label{sec:gw}

The contribution of cosmic string loops to the stochastic background
of gravitational waves was recognized as one of the most significant
observational signatures of a network of strings, and over the years it
has been calculated by several groups with various different assumptions
\cite{Vilenkin:1981bx,Hogan:1984is,Vachaspati:1984gt,Accetta:1988bg,
  Bennett:1990ry,Caldwell:1991jj,Siemens:2006yp,DePies:2007bm,Olmez:2010bi,Sanidas:2012ee,Sanidas:2012tf,Binetruy:2012ze,Kuroyanagi:2012wm,
  Kuroyanagi:2012jf}.

Having obtained a robust description of the distribution of
loops from our simulations we can now use it to update the
calculation of the spectrum of energy in gravitational waves, $\Omega_{\rm
  gw}(\ln f)$, with $\Omega_{\rm gw}(\ln f)d\ln f$ being the fraction
of the critical density in gravitational waves whose frequencies lie
between $f$ and $f +  f d\ln f$.  We define the
gravitational wave energy per (physical) volume per unit frequency
$\rho_{\rm gw}(t,f)$, and and then divide by the critical energy
density to obtain
\beq
\Omega_{\rm gw}(\ln f) = \frac{8\pi G}{3 H_0^2} f \rho_{\rm gw}(t_0,f).
\eeq

We let ${\cal P}_{\rm gw}(t,f)df$ be the power per physical volume
flowing into gravitational waves of frequencies between $f$ and $f +
df$.  Then $\rho_{\rm gw}$ is just the time integral of ${\cal P}_{\rm
  gw}$, taking into account the $a^{-4}$ scaling of radiation energy
densities and the redshifting of frequency,
\beq
\rho_{\rm gw}(t,f) = \int_0^{t}dt' \frac{a^4(t')}{a^4(t)}{\cal P}_{\rm
  gw}(t',F')\frac{\partial F'}{\partial f}
=\int_0^{t} dt' \frac{a^3(t')}{a^3(t)}{\cal P}_{\rm gw}\left(t',\frac{a(t)}{a(t')}f\right),
\eeq
where $F' =\frac{a}{a'}f$.

The power per volume ${\cal P}_{\rm gw}$ is given by the comoving number density of loops $n(t,m)$, and the power $P$ radiated by each loop,
\beq
{\cal P}_{\rm gw}(t,f) = \int dm\, \frac{n(t,m)}{a^3(t)} P(m,f).
\eeq
Assuming most loops develop and maintain at least one cusp per oscillation for much of their evolution, the power radiated by a slowly moving loop of mass $m$ is approximately
\beq
P(m,f) = \frac{\Gamma G\mu^2}{\zeta(\tfrac{4}{3},j_\ast)} \sum_{j = 1}^{j_\ast} j^{-4/3}\delta\left(f - j\frac{2\mu}{m}\right),
\eeq
where the maximum allowed harmonic is $j_\ast$ (which we will assume to be infinite), $\zeta(\tfrac{4}{3},j_\ast) = \sum_{j=1}^{j_\ast}j^{-4/3} \approx 3.60$, and we have used the fact that the the period of the loop is $\frac{m}{2\mu}$.  The spectral index $q = 4/3$ represents the high frequency behavior of cusp emission.

Putting this together yields
\beq
\rho_{\rm gw}(t_0,f) &=&  \frac{\Gamma G\mu^2}{\zeta(\tfrac{4}{3},j_\ast)}\sum_{j = 1}^{j_\ast} j^{-4/3}\int_0^{t_0} dt\int dm\,  \frac{n(t,m)}{a^3(t_0)}\delta\bigg(\frac{a(t_0)}{a(t)}f - j\frac{2\mu}{m}\bigg),
\eeq
and so the fractional spectrum today is
\beq
\hspace{-4mm}\Omega_{\rm gw}(\ln f) 
&=& \frac{8\pi\Gamma G^2\mu^2f}{3H_0^2\zeta(\tfrac{4}{3},j_\ast)}\sum_{j = 1}^{j_\ast} j^{-4/3}\int_0^{t_0} dt \int dm\, \frac{n(t,m)}{a_0^3}\delta\bigg(\frac{a_0}{a(t)}f - j\frac{2\mu}{m}\bigg),\\
&=&\frac{8\pi\Gamma G^2\mu^2f}{3 H_0^2\zeta(\tfrac{4}{3},j_\ast)}\hspace{-1mm}\sum_{j = 1}^{j_\ast} j^{-4/3}\hspace{-1mm}\int_0^\infty\hspace{-3mm}\frac{dz}{H(z)(1+z)}\hspace{-1mm}\int\hspace{-1mm}dm\frac{n(t(z),m)}{a_0^3}\delta\bigg(\hspace{-1mm}(1+z)f - j\tfrac{2\mu}{m}\hspace{-1mm}\bigg),\\
&=&\frac{8\pi\Gamma G^2\mu^2}{3 H_0^2\zeta(\tfrac{4}{3},j_\ast)}\sum_{j = 1}^{j_\ast} j^{-4/3}\int_0^\infty \frac{dz}{H(z)(1+z)^3} \frac{n(t(z),\frac{2\mu j}{(1+z)f})}{a_0^3}\frac{2\mu j}{f} ,\label{eq:Omega-z}
\eeq
where we have used the fact that $dt = - dz/\left[H(z)(1+z)\right]$.
For $z\gtrsim 1$ the functions $H(z)$ and $t(z)$ are well approximated by the matter-plus-radiation solution
\beq
H_{m+r}(z) &=& H_0\sqrt{\Omega_{m,0}\left(1 + z\right)^3\left(1 + \frac{1+z}{1+\zeq} \right)} \label{eq:H-m+r-z}\\
t_{m+r}(z) &=&  \frac{2}{3 H_0\sqrt{\Omega_{m,0}}(1 + \zeq)^{3/2}}\left(2 + \sqrt{1 + \frac{1+\zeq}{1+z}}\left( \frac{1 + \zeq}{1+z} - 2\right) \right).\label{eq:t-m+r-z}
\eeq

Although we neglect the details of loop production in a matter-plus-radiation universe, the radiation era comoving loop number density is not significantly affected by the smooth transition to matter domination, certainly for the small ($\alpha \sim \Gamma G\mu$) loops which are important.  Thus for $t \leq \teq = t_{m+r}(\zeq)$, the comoving loop number density is well approximated by the pure radiation value of Eq.~(\ref{eq:approxnr}), converted to comoving units using Eq.~(\ref{eq:nofalpha}),
\beq
n_r(t(z),m) \approx \frac{a^3_r}{\mu d_{h,r}^4}n_r(\alpha_r),\label{eq:n-rad-z}
\eeq
where $\alpha_r = \frac{m}{\mu d_{h,r}}$, $d_{h,r} = 2 t(z)$, and $a_r = \sqrt{\frac{t(z)}{\teq}}\frac{a_0}{1+\zeq}\frac{2}{\sqrt{3 + 3/\sqrt{2}}}$.  The numerical factors in $a_r$ are chosen such that $a_r \to a_0/(1+z)$ for $z \gg \zeq$.

For $t > \teq$, the most numerous loops are those which survive from the radiation era,
\beq
n_{r\to m}(t(z),m) = n_r(\teq,m + \Gamma G\mu^2[t(z)-\teq]).\label{eq:n-radmat-z}
\eeq

The loops produced during the matter era are given by Eq.~(\ref{eq:transientnmat}) and
\beq
n_m(t(z),m) = \frac{a_m^3}{\mu d_{h,m}^4} n_m(t(z),\alpha_m),\label{eq:n-mat-z}
\eeq
where $\alpha_m = \frac{m}{\mu d_{h,m}}$, $d_{h,m} = 3 t(z)$, and $a_m = \left(\frac{t(z)}{\teq}\right)^{2/3}\frac{a_0}{1+\zeq}(2-\sqrt{2})^{2/3}$.  The numerical factors in $a_m$ are chosen such that $a_m \to a_0/(1+z)$ for $z \ll \zeq$.

Dark energy becomes important for $z \lesssim 1$.  Although the additional dilution is already taken into account since we use comoving number densities, we should now compute cosmic time $t(z)$ using the $\Lambda$-plus-matter solution
\beq
H_{\Lambda+m}(z) &=& H_0\sqrt{\Omega_\Lambda + \Omega_{m,0}(1 + z)^3},\label{eq:H-l+m-z}\\
t_{\Lambda+m}(z) &=& \frac{2\tanh^{-1}\left(H_0\sqrt{\Omega_\Lambda}/H_{\Lambda+m}(z) \right)}{3 H_0\sqrt{\Omega_\Lambda}} + C_9,\label{eq:t-l+m-z}
\eeq 
where $\Omega_\Lambda = 1-\Omega_{m,0}$, and $C_9$ is chosen to make $t_{\Lambda+m}(z) = t_{m+r}(z)$ at some matter dominated splicing redshift, say $z = 9$.

By using the matter-era simulated $n_m(\alpha)$, and just plugging $t(z) \to t_{\Lambda + m}(z)$ in Eq.~(\ref{eq:n-mat-z}) and below, we are making the assumption that
the comoving loop production function is unaffected by dark energy.  
This is certainly true for all but the largest loops, since production of small loops is always negligible.

Put another way, the vast majority of loops today were produced at $z
\gg 1$, when matter-era simulation data can be trusted.  The Boltzmann
equation, Eq.~(\ref{eq:boltzmann-solution}), is just
\beq
n(t,m) = n\big(t_i,m + \Gamma G\mu^2 (t - t_i)\big) + \int_{t_i}^t f\big(t',m + \Gamma G\mu^2 (t - t')\big)dt',
\eeq
which says that for any loop sizes whose recent production rate is
negligible (i.e., $f(t',M') = 0$), we simply take the previously
understood comoving number density at time $t_i$, and age it by $t -
t_i$.  It is irrelevant whether we use $f_m$ or the true
$f_{\Lambda+m}$, since they both vanish for the vast majority of loops
which exist for $0\leq z \lesssim 1$.  Hence using $n_m(t,m)$ today is
justified provided we don't care about loops larger than about 2\% of
the size of the observable universe.  Note that this is only true in
comoving units, since scaling units assume the wrong horizon size, and
physical units neglect the extra dilution due to dark energy.  We can
always convert to physical units using scale factor $a_0/(1+z)$,
rather than $a_m(t_{\Lambda+m}(z))$.

\begin{figure}
   \centering
   \hspace{0cm}  
    \includegraphics[width=6in]{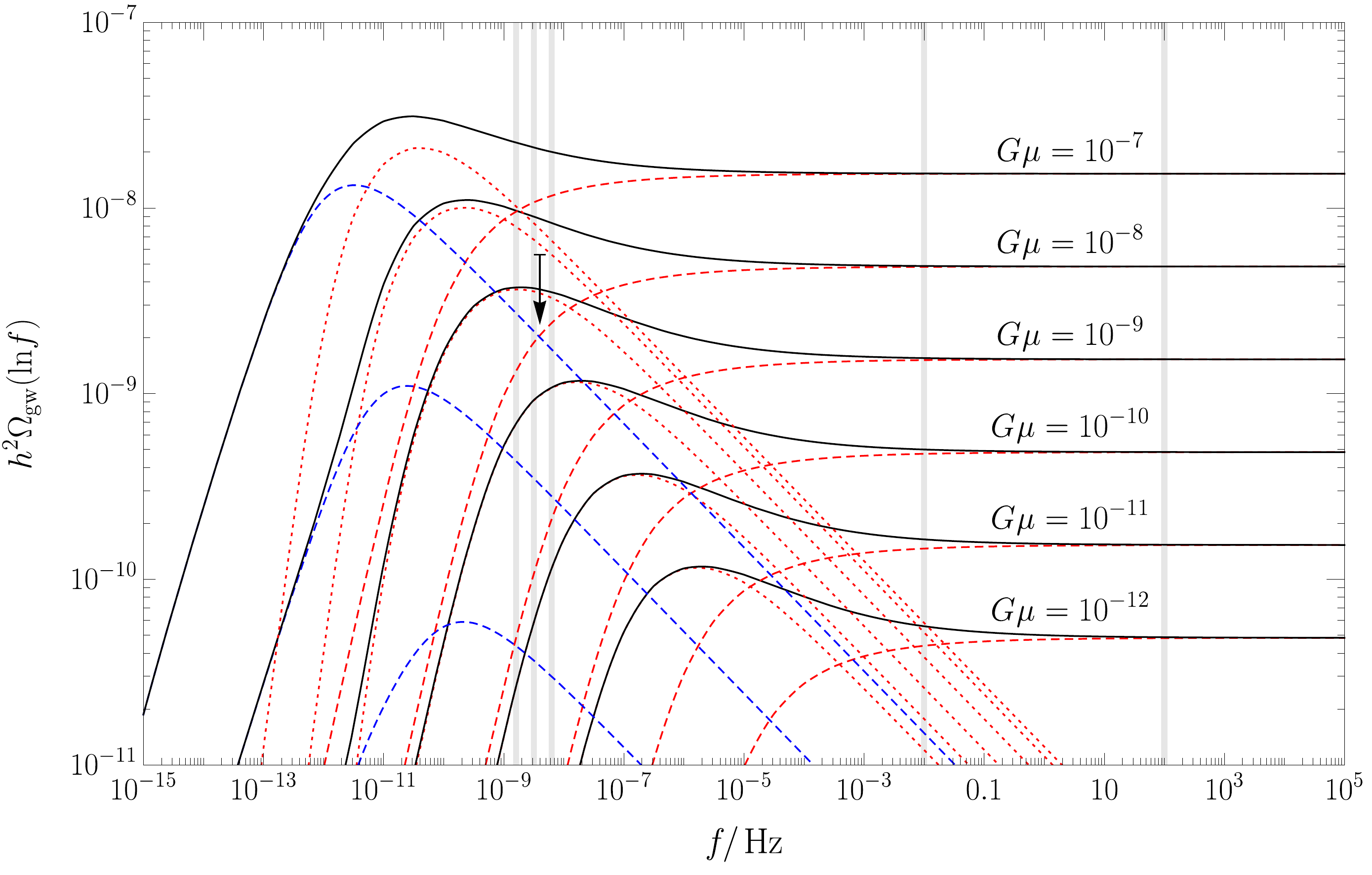} 
   \caption{The normalized spectrum of gravitational waves for various values of string tension.  The red dashed lines show the contribution from loops radiating during the radiation era, the red dotted lines represent the contribution from loops produced during the radiation era, but radiating during the matter era, and the blue dashed lines represent loops produced during the matter era.  In light gray from left to right, the 20-, 10-, and 5-year PTA,  eLISA \cite{Binetruy:2012ze, AmaroSeoane:2012je} and LIGO \cite{Abbott:2009rr} peak sensitivity frequencies are shown.}
   \label{fig:OmegaGW}
\end{figure}

Putting the comoving number densities,
Eqs.~(\ref{eq:n-rad-z}--\ref{eq:n-mat-z}), into Eq.~(\ref{eq:Omega-z}),
and using the analytic approximations for the Hubble rate and proper
time, Eqs.~(\ref{eq:H-m+r-z}--\ref{eq:t-m+r-z}) and
Eqs.~(\ref{eq:H-l+m-z}--\ref{eq:t-l+m-z}) for redshifts above and below
$z=9$, respectively, we numerically compute the stochastic
gravitational wave background for various values of $G\mu$, assuming
$\Gamma = 50$, shown in Fig.~\ref{fig:OmegaGW}.

The high frequency behavior is independent of the spectral index $q = 4/3$, and can be written analytically for $f \gg (1+\zeq)/(\Gamma G\mu \teq)$ as
\beq
\Omega_{\rm gw}(\ln f) &\to& \frac{8\pi\Gamma G^2\mu^2\Omega_{m,0}}{3(1+\zeq)}\int_0^1 n_r(\alpha)d\alpha,\\
&\approx& 8.14\frac{\Omega_{m,0}}{1+\zeq}\sqrt{\frac{G\mu}{\Gamma}},\\
&\approx & 7.7\times 10^{-4}\sqrt{\frac{G\mu}{\Gamma}},
\eeq
where we are using the best-fit cosmological parameters from Planck + WMAP polarization \cite{Ade:2013zuv}, $\zeq = 3403$, $\Omega_{m,0} = 0.3183$, and $H_0 = 2.171\times 10^{-18\,}{\rm Hz}$, i.e., $h = 0.6704$.

\begin{figure}
\centering
\includegraphics[width=5in]{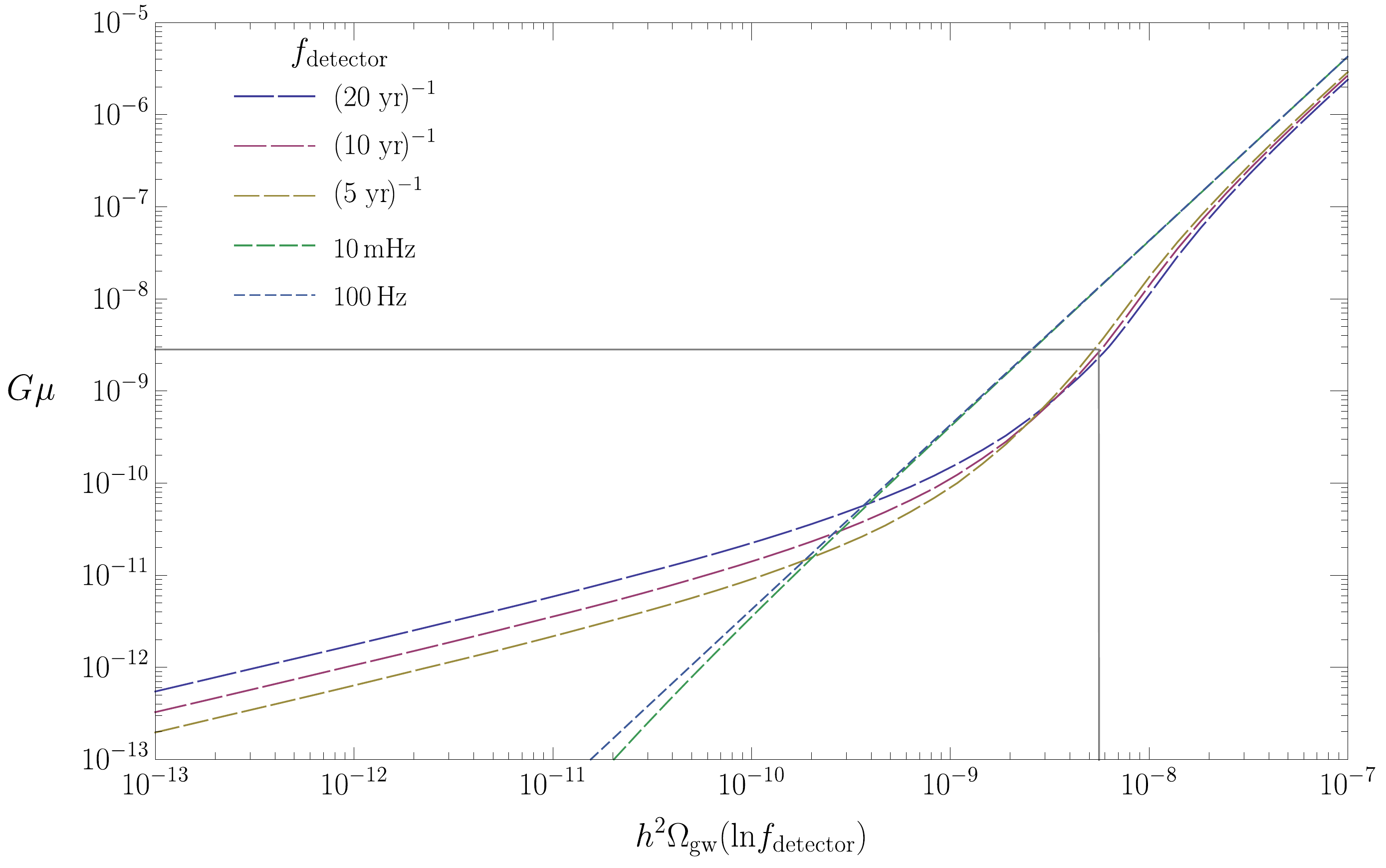}
\caption{Constraints on $G\mu$ for five detector peak sensitivity frequencies.  The horizontal axis is the upper bound on $h^2\Omega_{\rm gw}$ provided by a particular experiment, and the vertical axis represents the corresponding upper bound on the cosmic string tension.  The five frequencies plotted are shown in the legend, with PTA detector frequencies given by the inverse of the duration of the experiment.  The gray solid lines indicate the bound of Eq.~(\ref{eq:bound}). }
\label{fig:PTA-Gmu}
\end{figure}

In Fig.~\ref{fig:PTA-Gmu} we graph the string tension $G\mu$
vs.\ the normalized power in gravitational waves for five
characteristic detector frequencies.  Currently, the most stringent
limits on $G\mu$ come from pulsar timing arrays
(e.g. \cite{vanHaasteren:2011ni}, \cite{Demorest:2012bv}).  Following
the analysis of Ref.~\cite{Sanidas:2012ee} on the data of
Ref.~\cite{vanHaasteren:2011ni}, we use the 95\% confidence limit
$h^2\Omega_{\rm gw}(f = 4.0\times10^{-9}{\rm Hz}) \leq 5.6\times
10^{-9}$, which using our loop distribution provides the bound on
tension
\beq\label{eq:bound}
G\mu \leq 2.8\times 10^{-9},
\eeq
as shown in Fig.~\ref{fig:PTA-Gmu}.  Notice that this bound
is consistent with the range of tensions expected from cosmic superstrings, which are produced after brane inflation
\cite{Firouzjahi:2005dh}.  This bound should not be taken as
definitive, since we have neglected several effects, including changes in the number of
relativistic degrees of freedom at early times, the actual spectrum of
gravitational emission from realistic loops, the possibility that some
energy is in rare bursts that we would not have observed \cite{Regimbau:2011bm}, and the
possible fragmentation of loops after significant gravitational backreaction.  We
intend to do a more careful analysis of the gravitational wave signature of loops
including all these effects.

We can compare this bound with the results of Ref.~\cite{Sanidas:2012ee}, which found $G\mu < 8.8\times 10^{-11}$ for the case
of all loops being produced with scaling energy $x = 0.05$.  We believe the order of magnitude
discrepancy is primarily due to the fact that only about $10\%$ of power actually flows into the largest loops, as discussed 
below Eq.~(\ref{eq:nrnum}).  A precise comparison is difficult, since both our loop sizes and velocities differ from models they
considered.

\section{Conclusions}

By extrapolation from loop production found in simulations, we give
the distribution of loops to be found in the universe at any given time.
While some uncertainties remain in the late-time behavior of the loop
production function, these make no significant contribution to the loop number density.
The numerical value for the loop number density arising from simulations over the past decade is now rather well agreed upon,
with numerical coefficients roughly of order unity \cite{Ringeval:2005kr,Vanchurin:2005pa,Olum:2006ix,BlancoPillado:2011dq}.

In the radiation era, in scaling units we find to good approximation
\be
\nr(\alpha) \approx \frac{0.52}{\left(\alpha + \Gamma G\mu/2\right)^{5/2}},
\ee
up to a cutoff at $\alpha \approx 0.05$.  In physical units, this
gives
\be\label{eq:nrphys}
\frac{\nr(t,l)}{a^3(t)}\approx \frac{0.18}{t^{3/2}(l + \Gamma G\mu t)^{5/2}},
\ee
up to a cutoff at $l \approx 0.1t$.  While the functional form agrees with Eq.~(10.1.12) of
Ref.~\cite{Alexbook}, it differs by a significant numerical factor.

In the matter era, we have 
\be
\nm(\alpha) \approx \frac{2.4 - 5.7\alpha^{0.31}}{\left(\alpha + \Gamma G\mu/3\right)^2}
\ee
for $\alpha<0.06$, or in physical units
\be\label{eq:nmphys}
\frac{\nm(t,l)}{a^3(t)} \approx \frac{0.27 - 0.45(l/t)^{0.31}}{t^2(l + \Gamma G\mu t)^2},
\ee
for $l < 0.18t$.  (We ignore the small loop cutoff of Eq.~(\ref{eq:alphamax}), since 
radiation era loops dominate regardless.) For $l\ll t$, the second term in the numerator can
be ignored, and then Eq.~(\ref{eq:nmphys}) agrees with Eq.~(10.1.17)
of Ref.~\cite{Alexbook} except for numerical factors.

Most importantly in the matter era, there will be loops remaining from the
radiation era, given by
\be
\nr(t>\teq,\alpha)\approx
\frac{0.94t^{1/2}}{\teq^{1/2} (\alpha+\Gamma G \mu/3)^{5/2}}
\ee
for $\alpha < 0.03(\teq/t)-\Gamma G \mu/3$, or
\be
\frac{\nr(t>\teq,l)}{a^3(t)}\approx
\frac{0.18\teq^{1/2}}{t^2 (l+\Gamma G \mu t)^{5/2}}
\ee
for $l < 0.09\teq-\Gamma G \mu t$.  This agrees with Eq.~(10.1.20) of Ref.~\cite{Alexbook},
except for a significant numerical factor, and the effects of
evaporation.

The above calculations are for a pure matter era (with a sudden
transition from a pure radiation era at $\teq$).  Embedding these
results into a realistic cosmology is done in Sec.~\ref{sec:gw}.  The
calculation is simpler if we are only interested in the loop number
density today, which we can compute in a way which depends only on the
present content and age of the universe, as follows.

First consider relics from the radiation era.  Let $t_r$ and $z_r$ be
some time and corresponding redshift in the radiation era.  The loop
density at $t_r$ is given by Eq.~(\ref{eq:nrphys}).  The density of
the same loops today is just given by dilution,
\be
\frac{n_r(t_0,l)}{a_0^3} 
= \frac{0.18}{(1+z_r)^3 t_r^{3/2} (l+\Gamma G\mu t_0)^{5/2}}
\ee
for $l < 0.09\teq-\Gamma G \mu t_0$.  In the radiation era, we have
\be
\frac{1}{4t_r^2} = H^2(t_r) = H_0^2 \Omega_{r,0}(1+z_r)^4
\ee
and thus
\be
\frac{n_r(t_0,l)}{a_0^3} 
= \frac{0.51 (H_0^2\Omega_{r,0})^{3/4}}{(l+\Gamma G\mu t_0)^{5/2}}
\ee
Here $\Omega_r$ refers to the radiation density that we would have
today with the present CMB temperature and neutrinos being massless,
i.e., the one that we could extrapolate backward into the radiation
era when neutrino masses did not matter.

Since
\be
H_0^2\Omega_{r,0} = \frac{8\pi G}{3}\rho_r(T_0),
\ee
$n_r(t_0,l)/a_0^3$ can be computed from the present-day CMB
temperature $T_0$ and the current age of the universe, $t_0$.

We can make a similar calculation for loops produced in the matter
era.  Let $t_m$ and $z_m$ be some time and redshift when the universe
was matter-dominated, before dark energy became important.  Use
Eq.~(\ref{eq:nmphys}) at $t_m$, and then advance to the present.  In
the matter era, we have
\be
\frac{4}{9t_m^2} = H^2(t_m) =  H_0^2\Omega_{m,0} (1+z_m)^3
\ee
and thus
\be\label{eq:nmt0}
\frac{n_m(t_0,l)}{a_0^3} \approx \frac{0.61 - 1.0(l/t_0)^{0.31}}{(l +
  \Gamma G\mu t_0)^2}H_0^2\Omega_{m,0},
\ee
The presence of $t_0$ in the numerator of Eq.~(\ref{eq:nmt0}) is
correct only in the approximation that loop production continues
unchanged during dark energy domination.  The opposite case, in which
loop production stops entirely, can be approximated by writing $t_m$
instead of $t_0$ in the numerator of Eq.~(\ref{eq:nmt0}), and setting
$t_m$ to the start of the dark energy era.  But the effect is small
and applies only to fairly large loops.  The density of the very
largest loops could be determined only by simulations of the matter to
dark energy transition, but such loops are so rare as to be of very
little observational interest.

\section*{Acknowledgments}

We thank Joe Polchinski, Christophe Ringeval, Xavier Siemens, Vitaly
Vanchurin and Alex Vilenkin for helpful discussions.  This work was
supported in part by the National Science Foundation under grants
0855447, 0903889, 1213888, and 1213930 and by the Spanish Ministry
of Science and Technology under the FPA2009-10612 and FPA2012-34456 grants.

\begin{appendix}
\section{Differential forms}\label{sec:appendix-forms}
\subsection{Introduction}

Quantities such as the density of loops are more elegantly described
using differential forms.  Thus instead of taking $n(\alpha)$ as our
fundamental quantity, we take the 1-form $n(\alpha)d\alpha$.  The
definition of this differential form is the object which, when integrated over a range of
$\alpha$, gives the number of loops whose sizes lie in that range.  This
avoids some awkwardness in defining the function $n(\alpha)$, and makes the transformation
between physical and scaling coordinates clear.

Since a differential form returns physical, coordinate-independent
values upon integration, it is the area under the curve (or surface)
that we use to represent it, rather than the height of the curve.  In
this regard it is very much like a probability density function.  For
example, in a graph of $n(\alpha)d\alpha$ with $\alpha$ on a linear scale,
the area under the curve $n(\alpha)$ faithfully represents the
integral $\int n(\alpha) d\alpha$.  But when we graph the same
differential form with $\alpha$ on a logarithmic scale, the horizontal
measure is now $d\log\alpha = \frac{d\alpha}{\alpha}$.  Thus we should
plot $\alpha n(\alpha)$, since the area under this curve is $\int
\alpha n(\alpha) d\log \alpha$, which is equal to the intended $\int
n(\alpha) d\alpha$.  This is the reason for the extra power of
$\alpha$ that appears in almost all of our figures.  When graphing a
2-form such as $\alpha^{3/2}f_r(\alpha,p)d\alpha\wedge dp$, shown in
Fig.~\ref{fig:a3-2fcontourrad}, we include both an extra power of
$\alpha$ and $p$, since both axes are logarithmic.

Our conjecture that the loop production power is a scaling quantity (even in the absence of gravitational back-reaction)
can be phrased in terms of the 1-form $x f(x) dx$, where $x$ is the scaling energy of the loop. 
Since the integral of a 1-form is a scalar function, we can define scaling of the loop production function to mean the integral ${\cal P}(t,x) = \int_0^x x' f(t,x')dx'$ converges to a time-independent smooth function ${\cal P}(x)$
that vanishes at $x=0$.  This means that no power flows into loops of size set by the initial conditions, or any size sufficiently far below the horizon scale.

\subsection{The Boltzmann equation}

We show how to compute the distribution of loops from their production
rate, in the language of differential forms.  This is the purpose of a
Boltzmann equation (see e.g., Refs.~\cite{Copeland:1998na,
  Leblond:2009fq, Peter:2013jj}), which describes the
(non)conservation of some current in phase space.  We will consider
the phase space to be spanned by the loop's mass and
momentum per unit mass, and so the observable quantity will be the comoving
number density, given by the two-form $n(t,m,p)\, dm\wedge dp$, which
describes the number of loops per comoving volume at time $t$ with
mass between $m$ and $m + dm$, and with momentum between $p$ and $p +
dp$.

Let us first imagine that all loops have the same mass $M(t)$ and
momentum $P(t)$, where these functions smoothly vary with time.
Further, if the loops are neither created nor destroyed, their
constant comoving number density $n$ is described by the conserved
current
\be\label{delta-function-J}
{\mathcal J} = n \delta(m - M(t))(dm - \dot{M}dt)\wedge \delta(p - P(t))(dp - \dot{P}dt),
\ee
where $\delta$ is the Dirac delta function, and $\dot{M}(t) = \frac{dM(t)}{dt}$, etc.
Notice that this current can be written 
\be
{\mathcal J} = nd\Theta(m - M(t))\wedge d\Theta(p - P(t)),
\ee
where $\Theta$ is the step function.  This form of $\mathcal J$ makes it clear that its contour surfaces are parallel
to $(M,P)$, i.e., the flow generated by $\dot{M}$ and $\dot{P}$.  Since $d^2 \equiv 0$, the current is indeed conserved:
\be
d{\mathcal J} = 0.
\ee

We can generalize the above current to have support on more than a single value of $m$ and $p$ by adding together many parallel currents.
We measure the number density on surfaces of constant $t$, so we should define $n(t,m,p)\,dm\wedge dp$ as the pullback of ${\mathcal J}$
onto this surface.  (In the above case, $n(t,m,p) = n\delta(m-M(t)) \delta(p-P(t))$.)  Now, the integral curves $M$ and $P$ will have two sets of arguments.  On the one hand,
$M = M(t')$, since a given loop's mass depends only on time.  But each integral curve depends only on its starting value, so $M = M(t';t,m,p)$, with $M(t;t,m,p) = m$.
Notice $\dot M(t,m,p) = \left.\frac{d}{dt'}M(t';t,m,p)\right|_{t' = t}$.

Let us consider a more general two-form $\mathcal J$ which is not necessarily conserved:
\be\label{boltzmann-abstract}
d\mathcal J = \mathcal F,
\ee
where $\mathcal F = f(t,m,p)\,dt\wedge dm\wedge
dp$ is an arbitrary three-form describing the number of loops
per comoving volume that are produced between times $t$ and $t + dt$
which have mass between $m$ and $m + dm$ and momentum between $p$ and
$p + dp$.  This is the abstract form of the Boltzmann equation.
Even the most general $\mathcal J$ we can consider is not completely
arbitrary, but describes a current which flows along the integral
curves of the vector field generated by $\dot{M}(t,m,p)$
and $\dot{P}(t,m,p)$, which means we can relate it to $n(t,m,p)$ by
\be
\mathcal J = n(t,m,p)\left(dm - \dot{M}(t,m,p)dt\right)\wedge \left(dp - \dot{P}(t,m,p) dt\right),
\ee
i.e., this is the unique two-form whose pullback vanishes
on surfaces parallel\footnote{In other words, ${\cal J}_{\mu \nu}{\cal V}^{\nu} = 0$, where ${\cal V} = \frac{\partial}{\partial t} + \dot M \frac{\partial}{\partial m} + \dot P \frac{\partial}{\partial p}$.}
 to the flow, and whose pullback on constant time surfaces is given by $n(t,m,p)\,dm\wedge dp$.  Thus we can rewrite the Boltzmann equation (\ref{boltzmann-abstract}) as
\be
d\left[n\,\left(dm - \dot{M}dt\right)\wedge \left(dp - \dot{P} dt\right)\right] = f\,dt\wedge dm\wedge dp,
\ee
from which we can derive the more familiar form\footnote{Notice since
  $n$ is not a scalar, the following is \emph{not} correct: $f =
  \frac{d}{dt}n = \frac{\partial}{\partial t}n + \dot M
  \frac{\partial}{\partial m}n + \dot{P} \frac{\partial}{\partial
    p}n$.  The missing term $n\frac{\partial}{\partial p} \dot P$ is
  important; Even for $f=0$, as loop momenta redshift, the peak value
  of $n$ grows to compensate its shrinking support.}
\be
\frac{\partial}{\partial t} n(t,m,p) + \frac{\partial}{\partial
  m}\Big(\dot{M}(t,m,p) \, n(t,m,p)\Big)
+\frac{\partial}{\partial p}\Big(\dot{P}(t,m,p)\,n(t,m,p)\Big) = f(t,m,p).\;\;
\ee
This form of the Boltzmann equation would be useful for determining $f$, given $n$.
We are interested in finding $n(t,m,p)$ for a given $f(t,m,p)$, which can be done using Stokes' theorem, which says
\be\label{eq:boltzmann-integral}
\int_B\mathcal F = \int_B d\mathcal J = \int_{\partial B} \mathcal J,
\ee 
where the last integral will contain the term $n(t,m,p)dm\wedge dp$ if we choose $B$ properly.
Thus we will find $n(t,m,p)$ by integrating the abstract Boltzmann equation, but along a specially chosen infinitesimally narrow volume.
Imagine drawing a narrow three dimensional pill box $B$ in extended phase space which begins at some initial time $t_i$, and ends at the time we are interested in, $t$.  The pill box intersects the $t$-surface to form an infinitesimal parallelogram centered at $(m,p)$ with sides $dm$ and $dp$.  We assume the initial boundary data  $n(t_i,m,p)$ are known for some $t_i$.  Finally, we will choose the four long sides of the pill box to follow the flow in phase space generated by $\dot{M}$ and $\dot{P}$. By doing this, we have ensured that in Eq.~(\ref{eq:boltzmann-integral}), the integral of $\mathcal J$ over the boundary of the pill box will vanish on four out of six sides, and so the bulk integral of $\mathcal F$ over $B$ will give us $n(t,m,p)$ in terms of $f$ and the initial boundary data $n(t_i,m,p)$.  We illustrate $B$ in Fig.~\ref{fig:pillbox} below.
 \begin{figure}
   \centering
   \includegraphics[width=4.0in]{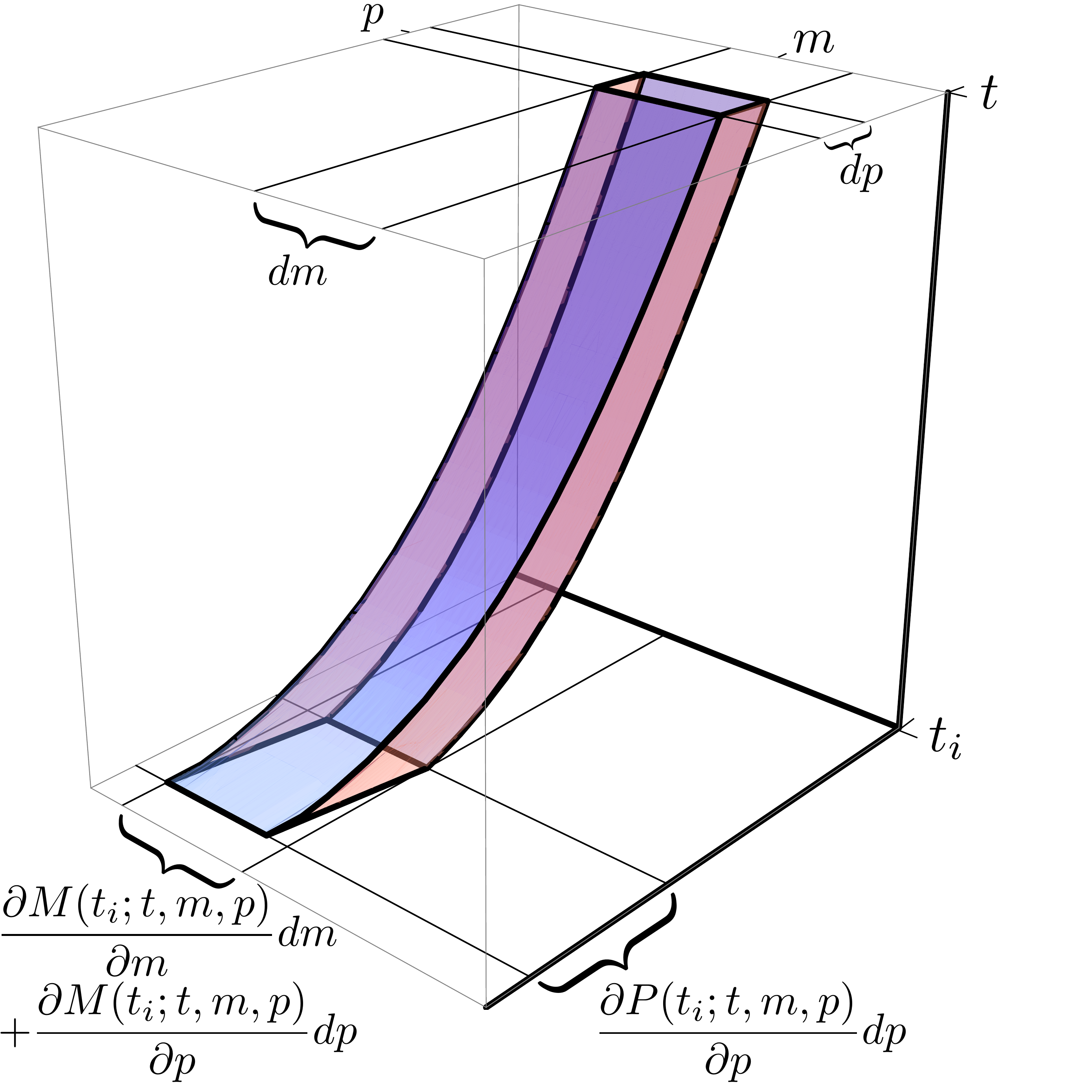} 
   \caption{The pill box $B$.  The boundary $\partial B$ consists of six faces, the four curved of which do not contribute to $\int \mathcal J$.  The top and bottom parallelograms do contribute and have areas given by $dm\wedge dp$ and $\left(\frac{\partial M_i}{\partial m}\frac{\partial P_i}{\partial p} - \frac{\partial M_i}{\partial p}\frac{\partial P_i}{\partial m}\right)dm\wedge dp$, respectively.} 
   \label{fig:pillbox}
\end{figure}

\subsection{The flow $(M,P)$}
To use the Boltzmann equation, we must know what generates the flow,
i.e., given any values for the mass and momentum $m$, $p$ at time $t$,
we must know the rate at which they change, $\dot{M}(t,m,p)$,
$\dot{P}(t,m,p)$.  These functions define a vector field $\cal V$ on
extended phase space,
\ba
{\cal V} = \frac{\partial}{\partial t} + \dot M \frac{\partial}{\partial m} + \dot P \frac{\partial}{\partial p}.
\ea
We can characterize the associated integral curves with the functions
$M' = M(t';t,m,p)$ and $P' = P(t';t,m,p)$.  These determine the entire
trajectory as a function of $t'$, given the initial starting point
$(m,p)$ at time $t$.  We solve an autonomous system of ordinary
differential equations to get from the generators $\dot{M}$ and
$\dot{P}$ to the curves $M'$ and $P'$, and so the solution is
unique\footnote{See \cite{Nakahara:book} for a review of flows.}.

\subsection{The solution}
We are now ready to solve for $n(t,m,p)$ by integrating the Boltzmann equation along the narrow pill box $B$.  The top face of $B$ is denoted $B\cap t$ and is a tiny parallelogram of area $dm\wedge dp$.  The bottom face is a parallelogram denoted $B\cap t_i$ and has area given by the Jacobian determinant 
\ba
\left|\frac{\partial \left(M_i,P_i\right)}{\partial\left(m,p\right)}\right|dm \wedge dp = \left(\frac{\partial M_i}{\partial m}\frac{\partial P_i}{\partial p} - \frac{\partial M_i}{\partial p}\frac{\partial P_i}{\partial m}\right)dm \wedge dp.  
\ea
The remaining four faces contribute nothing in Eq.~(\ref{eq:boltzmann-integral}).  Using Stokes' theorem,
\ba
n(t,m,p)dm\wedge dp &=& \int_{B\cap t}\mathcal
J,\\ &=&\int_B\mathcal F \;+\; \int_{B\cap t_i}\mathcal
J,\\ &=&\left[\int_{t_i}^{t} f\left(t',
  M(t';t,m,p),P(t';t,m,p)\right)\left|\frac{\partial
    \left(M(t'),P(t')\right)}{\partial \left(m,p\right)}\right|dt'
  \right. \\ &&\left.+ \; n(t_i,M(t_i;t,m,p),P(t_i;t,m,p))
  \left|\frac{\partial \left(M(t_i),P(t_i)\right)}{\partial
    \left(m,p\right)}\right| \right] dm\wedge dp \nonumber,
\ea
or
\be\label{eq:boltzmann-solution}
n(t,m,p) = n(t_i,M_i,P_i)\left|\frac{\partial \left(M_i,P_i\right)}{\partial \left(m,p\right)}\right| \;+\;
\int_{t_i}^t f(t',M',P') \left|\frac{\partial \left(M',P'\right)}{\partial \left(m,p\right)}\right|  dt' .
\ee
Notice we have defined the integral curves $M(t')$ and $P(t')$ by integrating
$\dot{M}$ and $\dot{P}$ from the final surface $t$ backward as a
function of $t'$.  This enabled us to keep explicit dependence on $m$
and $p$.  We must account for the fact that the pill box
cross-sectional area changes by a factor of the Jacobian determinant,
as shown in Fig.~\ref{fig:pillbox}.  In our approximation, because the
momentum of a loop redshifts without any regard for the value of the
mass, the Jacobian determinant has one term, $\frac{\partial
  M}{\partial m}\frac{\partial P}{\partial p}$.

\section{The exact flow}\label{sec:exact-flow}

Here we give the solution to the flow $(M,P)$ for arbitrary loop
momentum per unit mass $p$.  The time derivatives which generate the
flow are given by
\beq
\dot M &=& -\Gamma G\mu^2/\gamma,\\
\dot P &=& - \frac{\dot{a}}{a} P,
\eeq
where $\gamma = 1/\sqrt{1-v^2} = \sqrt{p^2 +1}$ is the time-dilation factor. 

Changing to redshift as a time variable, the solution for $P$ is just
\beq
P'=P(z';z,m,p) =p\frac{1 + z'}{1+z},
\eeq
as before.   The loop mass obeys
\beq\label{eq:exactM}
M'=M(z';z,m,p) =-\Gamma G\mu^2\int\frac{1}{\gamma'}\frac{dt'}{dz'}dz' = \Gamma G\mu^2\int\frac{1}{(1+z')H(z')\gamma'}dz' ,
\eeq
where
\be
\gamma' &=&  \sqrt{P'^2+1},\label{eq:gammasol}.
\ee
If we neglect dark energy, we can write
\be
H(z) = H_0\sqrt{\Omega_{m,0}\left(1 + z\right)^3\left(1 + \frac{1+z}{1+\zeq} \right)},\label{eq:Hofz}
\ea
which makes it possible to perform the integral in
Eq.~(\ref{eq:exactM}) in closed form, but the result in terms of
elliptic integrals is not very enlightening.

\section{Tiny, recent loops}\label{sec:appendix-tiny}

We assumed above that long strings are smoothed by gravitational back
reaction at scales $\Gamma G\mu t$, and thus that no loops are
produced at time $t$ with $\alpha\ll\Gamma G\mu t$.  If this is not the
case, we can have a different situation.  Loops formed recently with
size $\alpha>\Gamma G \mu$ are always dominated by loops formed long
ago, when the string network was denser.  But when $\alpha\ll\Gamma
G\mu$, the possible formation time for a loop that currently
has size $\alpha$ no longer decreases with $\alpha$, because loops
from too long ago have evaporated.  This raises the possibility, which
we investigate in this appendix, that such tiny loops are mainly of
recent origin.

Thus we consider a second population of loops created with $\alpha \ll \Gamma
G\mu$.  Such loops live for much less than a Hubble time, so for them
we can neglect redshifting and increase in $t$ from formation to
observation.  Thus we set $a(t') = a(t)$, $\dh(t') = \dh(t)$, and $P'
= p$ in Eq.~(\ref{eq:nrm1}) to get
\be
\nrecent(\alpha,p)=\int_0^t\frac{1}{\dh}f\left(\alpha+\Gamma
G \mu (t-t')/\gamma, p\right) dt'
\ee
where $\gamma = \sqrt{p^2+1} \sim p$ is the Lorentz boost of a
loop with momentum $p$.  Changing variables, we find
\be
\nrecent(\alpha,p) = \frac{\gamma}{\Gamma G \mu} \int_\alpha^\infty f(\alpha',p)
d\alpha'
\ee

If we are not concerned with loop velocities, we can approximate
\be\label{eq:nrecent}
\nrecent(\alpha) = \frac{1}{\Gamma G \mu} \int_\alpha^\infty
\gamma(\alpha') f(\alpha') d\alpha'
\ee
where $\gamma(\alpha')$ denotes the average boost with which loops of
mass $\alpha'$ are emitted.  The production rate $f(\alpha')$
increases more quickly than $1/\alpha'$ as $\alpha'$ decreases, until
reaching some cutoff due to gravitational smoothing of the long-string
network.  The cutoff is given by the shortest-wavelength wiggles that
have survived gravitational damping.  If this wavelength were $\Gamma
G \mu t$, the population of recent loops would be of no consequence.
But Ref.~\cite{Siemens:2002dj} argued that short-wavelength
structures on long strings will be damped slowly, because they are
only able to interact with wiggles of similar wavelengths, so we will
consider a cutoff
\be\label{eq:delta}
\lambda_{\rm min} \sim (\Gamma G\mu)^\delta t
\ee
which gives a limit
\be
\xmin \sim (\Gamma G\mu)^\delta
\ee

According to Ref.~\cite{Siemens:2002dj}, the smallest scales depend on
the falloff of the amplitude of structures on the string with
decreasing wavelength $\lambda$.  But since strings have kinks, the
amplitude-to-wavelength ratio of the Fourier components must fall no
more rapidly than $\lambda$, which
implies\footnote{Ref.~\cite{Siemens:2002dj} gave a model with
  $\delta=5/2$ in the matter era, but this model assumed no
  intercommutations and  thus no kinks.} $\delta \ge 3/2$.

Below $\alpha = \amin$, $\nrecent$ is a constant
\be
\label{eq:nrecent0}
\nrecent(0) = \frac{1}{\Gamma G \mu} \int_\amin^\infty
\gamma(\alpha') f(\alpha')
d\alpha'
\ee
If we take a model in which the formation of loops with $\xmin < x \ll
\Gamma G \mu$ is given by
\be
f(x) = c x^{-\beta}
\ee
and all loops of size $x$ have boost
\be
\gamma(x) = c' x^{-\kappa}
\ee
and thus $\alpha(x) = x^{1+\kappa}/c'$.
Then
\ba
\nrecent(0) &=& \frac{1}{\Gamma G \mu} \int_\xmin^\infty
\gamma(x) f(x) dx'
=\frac{cc'}{\Gamma G \mu (\beta+\kappa - 1)\xmin^{\beta+\kappa-1}}\\
&=& \frac{cc'}{(\beta+\kappa - 1)(\Gamma G \mu)^{1+(\beta+\kappa-1)\delta}}
\ea

For small $\alpha$, Eq.~(\ref{eq:n}) also gives a constant $n(\alpha)$
as $\alpha\to0$.  The ratio of recently produced tiny loops to old
tiny loops is
\be
\frac{\nrecent(0)}{n(0)} = 
\frac{cc'}{(1-\nu)(\beta+\kappa - 1)}\frac{(\Gamma G
  \mu)^{3-3\nu-(\beta+\kappa-1)\delta}}
{\int_0^\infty {\alpha'}^{3-3\nu}f\left(\alpha'\right)d\alpha'}
\ee
In the radiation era, fits to simulation data give
\ba
f(x) &\approx& 1.64 x^{-1.97}\\
\gamma(x) &\approx& 0.50 x^{-0.21}\,
\ea
which with $\delta=3/2$ give
\be
\frac{\nrecent(0)}{\nr(0)} = 1.64(\Gamma G \mu)^{-0.27}
\ee
In the matter era, we find
\ba
f(x) &\approx& 3.40 x^{-1.78}\\
\gamma(x) &\approx& 0.40 x^{-0.30}\,
\ea
which give
\be
\frac{\nrecent(0)}{\nm(0)} = 4.4(\Gamma G \mu)^{-0.62}
\ee
Since $\Gamma G \mu\ll 1$, the negative exponents imply that most tiny
loops are recently produced.

The total number density of recently-produced loops with $x < \Gamma G
\mu$ is
\be
\nrecent = \int_0^{\Gamma G \mu} dx\, x f(x)
= \frac{c (\Gamma G \mu)^{1-\beta}}{2-\beta}.
\ee
In the radiation era,
\be
\nrecent \approx 54 (\Gamma G \mu)^{-0.97},
\ee
which is much smaller than the density of old loops given by
Eq.~(\ref{eq:totalr}).  In the matter era,
\be
\nrecent \approx 16 (\Gamma G \mu)^{-0.78}.
\ee
For $\Gamma G \mu<10^{-6}$, this is no more than 1/8 of the number of
old loops given by Eq.~(\ref{eq:totalm}).  If one considers loop
energies, the recent loops make an even smaller contribution.

\end{appendix}

\bibliography{no-slac,boltzmann}

\end{document}